\documentclass[onecolumn,draftclsnofoot,12pt]{IEEEtran}
\usepackage{amssymb}
\usepackage{amsfonts}

\usepackage{enumerate}

\usepackage{enumerate}
\usepackage{amsmath,amsthm}
\usepackage{mathtools}
\usepackage{algorithm,algorithmic}
\usepackage{float}
\usepackage{hyperref}
\usepackage{color}
\usepackage{makeidx}
\usepackage{bbm}
\usepackage{graphicx}
\usepackage{lipsum}
\usepackage{soul}
\usepackage{tabularx}
\usepackage{dsfont}
\usepackage[table,xcdraw]{xcolor}

\usepackage{amsfonts}
\usepackage{times}
\usepackage{graphicx}
\usepackage{latexsym}
\usepackage{dsfont}
\usepackage{amssymb}
\usepackage{amsmath}
\usepackage{cite}
\usepackage{verbatim}
\usepackage{subfigure}

\newcommand{\tabref}[1]{{Table}~\ref{#1}}

\def\bb0{{\mathbb{0}}}

\def\bb{{\mathbf{b}}}

\def\bh{{\mathbf{h}}}

\def\bw{{\mathbf{w}}}

\def\b0{{\mathbf{0}}}

\def\sf0{{\mathsf{0}}}

\usepackage{epstopdf}

\newcommand{\sref}[1]{{Section}~\ref{#1}}
\newcommand{\fref}[1]{{Fig.}~\ref{#1}}

\newcommand{\tref}[1]{{Table}~\ref{#1}}

\DeclareMathOperator*{\argmax}{arg\,max}

\newcommand{\subto}{\operatorname{s.t.}}

\begin{document}
\title{Online Beam Learning with Interference Nulling for Millimeter Wave MIMO Systems}
\author{Yu Zhang, Tawfik Osman, and Ahmed Alkhateeb \thanks{Yu Zhang, Tawfik Osman, and Ahmed Alkhateeb are with Arizona State University (Email: y.zhang, tmosman, alkhateeb@asu.edu). This work is supported in part by the National Science Foundation under Grant No. 1923676 and by a U. S. Army research program under contract No. W911NF21C0015. Part of this work has been accepted in the Asilomar Conference on Signals, Systems, and Computers, 2022 \cite{Zhang2022Online}.}}
\maketitle

\begin{abstract}
Employing large antenna arrays is a key characteristic of millimeter wave (mmWave) and terahertz communication systems. Due to the hardware constraints and the lack of channel knowledge, codebook based beamforming/combining is normally adopted to achieve the desired array gain. However, most of the existing codebooks focus only on improving the gain of their target user, without taking interference into account. This can incur critical performance degradation in dense networks. In this paper, we propose a sample-efficient online reinforcement learning based beam pattern design algorithm that learns how to shape the beam pattern to null the interfering directions. The proposed approach does not require any explicit channel knowledge or any coordination with the interferers. Simulation results show that the developed solution is capable of learning well-shaped beam patterns that significantly suppress the interference while sacrificing tolerable beamforming/combing gain from the desired user. Furthermore, a hardware proof-of-concept prototype based on mmWave phased arrays is built and used to implement and evaluate the developed online beam learning solutions in realistic scenarios. The learned beam patterns, measured in an anechoic chamber, show the performance gains of the developed framework and highlight a promising machine learning based beam/codebook optimization direction for mmWave and terahertz systems.
\end{abstract}

\section{Introduction} \label{intro}

Millimeter wave (mmWave) and terahertz (THz) communication systems need to employ large antenna arrays to combat the severe path-loss and achieve sufficient receive signal power. Given the high cost of the mixed-signal components, these systems rely mainly on fully-analog or hybrid analog-digital architectures with much smaller number of RF chains compared to the number of antennas \cite{Alkhateeb2014MIMO,Heath2016,Molisch2017}.  These architectures, however, make it hard to explicitly estimate the wireless channels, which motivated these systems to rely on pre-defined beam codebooks for both initial access and data transmission \cite{Molisch2017,Zhang2022Reinforcement,Alrabeiah2020Neural,Heng2021}. Being pre-defined, however, those beams are normally designed in a way that focuses solely on improving the beamforming/combining gain from specific directions, without taking \textit{interference} into account. This leads to sub-optimal performance in dense deployments or in scenarios with intended interference/jamming. This urges the research for advanced analog/hybrid beam design approaches that are \textit{interference-aware}. Realizing that, however, is challenging because (i) these beams need to be designed online in the filed and (ii) without explicit channel knowledge which is hard to acquire for analog/hybrid architectures (especially for the interfering transmitters). With this motivation, this paper focuses on developing a beam learning framework that is able to learn interference nulling beam patterns while respecting the hardware and system operation constraints.

\subsection{Prior work}
Achieving effective interference suppression and management is a key research topic that attracted significant interest in various systems, such as  network MIMO \cite{Venkatesan2007,Yetis2010}, distributed antenna/coordinated multipoint (CoMP) systems  \cite{Choi2007,Gesbert2010,Irmer2011}, cell free massive MIMO \cite{Ngo2017,Interdonato2020,Demirhan2022}, and spectrum sharing and radar/communication coexistence \cite{Zhang2008Exploiting,Li2016Optimum,Qian2018}. The approaches developed in \cite{Venkatesan2007,Yetis2010,Choi2007,Gesbert2010,Irmer2011,Ngo2017,Interdonato2020,Zhang2008Exploiting,Li2016Optimum,Qian2018}, however,  considered conventional fully-digital MIMO transceiver architectures where the precoding and combining are implemented completely in the digital domain.
With the recent evolution towards high frequency bands (e.g., mmWave and terahertz) and the utilization of even larger antenna arrays, new transceiver architectures (e.g., fully analog and hybrid analog/digital architectures) are introduced to reduce the hardware cost and power consumption. These architectures rely on implementing the beamforming in the analog domain or distribute it between the analog and digital domains. This required revisiting the precoding design problem for single-user hybrid analog/digital architectures \cite{Alkhateeb2014,Ayach2014,Yu2016}, multi-user mmWave MIMO systems \cite{Alkhateeb2015Limited,Sohrabi2016,Zhan2021Interference}, mmWave full-duplex and relay systems \cite{Satyanarayana2019,Roberts2021,Zhu2020,Zhang2019On}, among others.

The existing precoding/beamforming approaches in \cite{Alkhateeb2014,Ayach2014,Yu2016,Alkhateeb2015Limited,Sohrabi2016,Zhan2021Interference,Satyanarayana2019,Roberts2021,Zhu2020,Zhang2019On}, however, have certain practical limitations: (i) Most of these solutions  mandate either full or partial channel state information in order to design the corresponding transmit/receive schemes. Acquiring this channel knowledge is challenging in realistic mmWave analog/hybrid architectures and is normally associated with large channel estimation and system coordination overhead (synchronization, information sharing among BSs or UEs, etc.). (ii) The existing approaches normally lack the capability/flexibility of performing online adjustment in order to account for imperfections in the hardware fabrication/calibration, errors in estimated channels, or  changes in the surrounding environment.
This motivates the development of \textbf{sample-efficient interference-aware beam learning approaches that do not require explicit channel knowledge of both the desired transmitter and the interfering transmitters, and do not any require coordination between the considered receiver and the interferers (which can not happen in practice).}

\subsection{Contribution}

Designing analog beam patterns that are aware of interference is an important problem and can be found in a variety of practical scenarios.
However, the hardware constraints associated with the fully analog transceiver architectures along with the scarcity of the accurate channel knowledge make the problem quite challenging.
In this paper, we propose a deep reinforcement learning based beam pattern design framework that can efficiently adapt the beam pattern to avoid interference sources while maximizing the beamforming/combining gain of the desired user. This is done by not requiring any explicit channel knowledge of the target user or the interferers, and by only relying on the power measurements. The proposed framework also respects the key hardware constraints of the analog/hybrid transceiver architectures, such as the quantized phase shifter and constant modulus constraints, making it a hardware compatible solution.
The main contributions of this paper can be summarized as follows:
\begin{itemize}
  \item \textbf{Designing a deep reinforcement learning based beam pattern design framework} that can learn how to shape nulls towards the directions of the interference. The developed solution relies only on receive power measurements and does not require any coordination between the considered receiver and the interfering transmitters. A 3GPP standard-compatible signal power estimation method is proposed to evaluate the signal reception performance under the presence of the always-on interfering transmitters.
  \item \textbf{Developing a model-based surrogate model assisted learning framework} that achieves higher sample efficiency by better leveraging the underlying signal model. The improvement on the sample efficiency has the potential of reducing the beam learning overhead as well as shortening the convergence time of the proposed solution. The proposed surrogate model also provides flexibility to the practical deployment in terms of enabling data sharing and cooperative (hence better learning) in  complex scenarios.
  \item \textbf{Extensively evaluating the proposed interference-aware beam learning solutions}  using numerical simulations. This provides a comprehensive assessment of the capability of the proposed learning approaches in nulling  interference without requiring any knowledge about the channel, array geometry, or user location.  The results also evaluate the potential performance and practical gains of the developed surrogate-model based learning framework.

  \item \textbf{Demonstrating the performance using a real-world proof-of-concept prototype}. To do that, we built a hardware prototype based on mmWave phased arrays and implemented the online beam learning solutions to run in real-time. The developed solutions are tested in the field under realistic conditions and the learned beam patterns are measured in an anechoic chamber. This provides important insights about how the proposed beam learning approaches perform in real-world scenarios and with practical hardware.
\end{itemize}

The real-world prototyping results show that the developed beam learning solution, which rely only on power measurements, can effectively suppress the interference with minimal impact on the desired signal gain. The results also show that proposed signal model-based surrogate model assisted learning framework significantly improves the sample efficiency, highlighting a promising interference-avoidance path for future mmWave and THz communication systems.

\section{System and Channel Models} \label{sec:System}

\begin{figure*}[t]
	\centering
	\includegraphics[width=.8\textwidth]{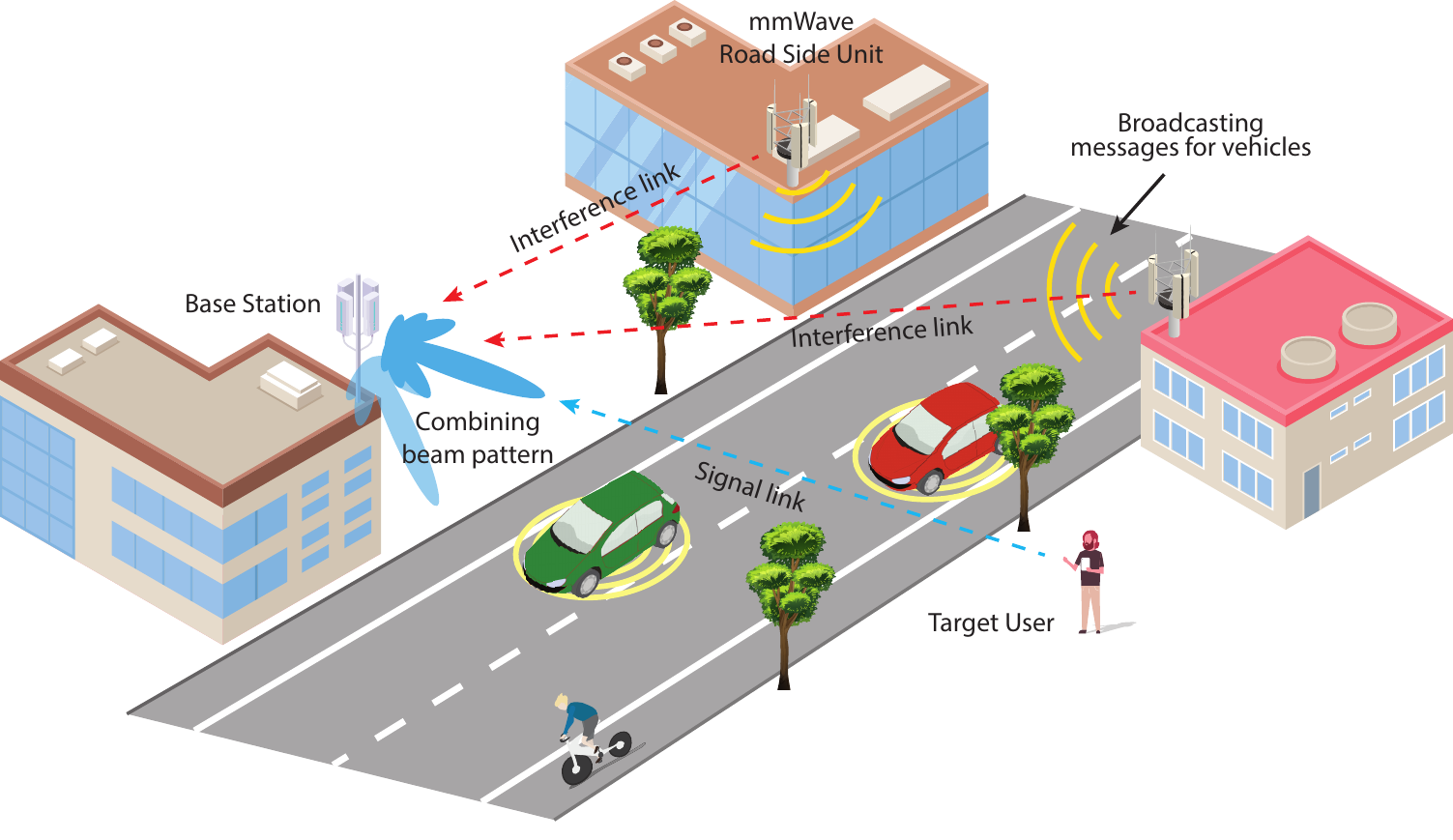}
	\caption{The considered uplink scenario where a mmWave base station, operating in a receive mode, is communicating with its target user under the presence of non-cooperative interference transmitters. This could be the case, for instance, where the mmWave road side units of a vehicular network are broadcasting traffic messages to the surrounding vehicles, which interferes the civilian data communication link, as depicted in the figure.}
	\label{fig:sys}
\end{figure*}

Our objective in this paper is to investigate the online design of interference-aware beam patterns, i.e., the online learning/design of beam patterns that achieve their original objectives while nulling the interference introduced by other sources in the environment.
To study this problem, we consider the communication system in \fref{fig:sys}, where a mmWave MIMO base station (BS), equipped with $M$ antennas, is communicating with a single-antenna user equipment (UE) in an {uplink mode}.
Moreover, we assume that there exist $K$ ($\ge 1$) non-cooperative interference transmitters\footnote{For ease of exposition, each interference transmitter is also assumed to have a single antenna. This means that the interference signals are being transmitted omni-directionally.} in the vicinity of the BS, operating at the same frequency bands and hence causing interference to the BS receiver.
Therefore, if the UE transmits a symbol $x\in\mathbb{C}$ to the BS, and the other $K$ interference transmitters also transmit symbols $x_k\in\mathbb{C}, k=1,\dots,K$ at the same time and frequency slot, such that all the transmitted symbols satisfy the same average power constraint, i.e., $\mathbb{E}[|x|^2]=P_x$ and $\mathbb{E}[|x_k|^2]=P_x, \forall k$, the received signal at the BS after combining can then be expressed as
\begin{equation}\label{rec}
  y = {\bf w}^H{\bf h} x + \sum_{k=1}^{K}{\bf w}^H{\bf h}_k x_k + {\bf w}^H{\bf n},
\end{equation}
where ${\bf h}\in\mathbb{C}^{M\times 1}$ is the channel between the BS and the UE, ${\bf h}_k\in\mathbb{C}^{M\times 1}$ is the channel between the BS and the $k$-th interference transmitter.
It is worth pointing out here that for clarity, we subsume the factors such as path-loss and transmission power into the channels.
${\bf n} \sim \mathcal{CN}(0, \sigma^2{\bf I})$ is the receive noise vector at the BS with $\sigma^2$ being the noise power and ${\bf w}\in\mathbb{C}^{M\times 1}$ is the combining vector used by the BS.
Furthermore, given the high cost and power consumption of the mixed-signal components, we consider a practical system where the BS has only one radio frequency (RF) chain\footnote{It is very important to note that the RF precoder in a system with hybrid architecture is normally constructed using pre-defined codebooks that have pre-determined beams. Therefore, the learned beams in this paper can be included in such codebooks and be used in the hybrid analog/digital architectures as well.} and employs analog-only beamforming/combining using a network of $r$-bit quantized phase shifters.
Therefore, the combining vector at the BS can be written as
\begin{equation}\label{Analog}
  {\bf w} = \frac{1}{\sqrt{M}}\left[ e^{j\theta_1}, e^{j\theta_2}, \dots, e^{j\theta_M} \right]^T,
\end{equation}
where each phase shift $\theta_m, \forall m=1,\dots,M$ is selected from a finite set $\boldsymbol{\Psi}$ with $2^r$ possible discrete values drawn uniformly from $(-\pi, \pi]$. The normalization factor $M^{-1/2}$ is to make sure the combiner has unit power, i.e., $\|\mathbf{w}\|_2^2=1$.

We adopt a  geometric channel model for the channel between BS and UE, as well as the interference channels between BS and any interfering transmitters.
Hence, the channel between BS and its served UE takes the following form (the channel between BS and any interference transmitter takes similar form)
\begin{equation}\label{ch}
  {\bf h} = \sum_{\ell=1}^{L} \alpha_{\ell}{\bf a}(\phi_{\ell}, \vartheta_{\ell}),
\end{equation}
where $L$ is the number of multi-paths.
Each path $\ell$ has a complex gain $\alpha_{\ell}$, which includes the  path-loss.
The angles $\phi_{\ell}$ and $\vartheta_{\ell}$ represent the $\ell$-th path's azimuth and elevation angles of arrival respectively, and ${\bf a}(\phi_{\ell}, \vartheta_{\ell})$ is the BS array response vector.
The exact expression of ${\bf a}(\phi_{\ell}, \vartheta_{\ell})$ depends on the array geometry and possible hardware impairments.

\section{Problem Formulation} \label{sec:special}

In this paper, we investigate the design of the analog combining/precoding that achieves interference awareness (i.e., attempts to address the interference) without explicitly knowing any channel state information.
Given the receive signal \eqref{rec} at the BS, the achievable rate of its target user can be written as
\begin{equation}\label{ach}
  R = \log_2\left(1 + \frac{|{\bf w}^H\mathbf{h}|^2P_x}{\sum_{k=1}^K|{\bf w}^H\mathbf{h}_{k}|^2P_x + \sigma^2}\right).
\end{equation}
The objective is to design the combining vector ${\bf w}$ such that the achievable rate of the target user, i.e., \eqref{ach}, can be maximized.
Given the monotonicity of the logarithm function, this is equivalent to maximize the SINR term in \eqref{ach}.
Therefore, the problem of designing interference-aware beam pattern can be cast as
\begin{align}\label{prob}
 {\bf w}^\star = \argmax\limits_{{\bf w}} & \hspace{2pt}  \frac{|{\bf w}^H\mathbf{h}|^2P_x}{\sum_{k=1}^K|{\bf w}^H\mathbf{h}_{k}|^2P_x + \sigma^2}, \\
 \subto  \hspace{2pt} &  w_m = \frac{1}{\sqrt{M}}e^{j\theta_{m}}, ~ \forall m=1, ..., M, \label{cons-1} \\
 & \theta_{m}\in\boldsymbol{\Psi}, ~ \forall m=1, ..., M, \label{cons-2}
\end{align}
where $w_m$ is the $m$-th element of the combining vector ${\bf w}$ \footnote{It is important to note that the proposed interference-aware beam  learning approach can be straightforwardly extended to learning a codebook with multiple beams by, for example, using the user clustering and assignment algorithm proposed in \cite{Zhang2022Reinforcement}.}.
The interference-aware beam pattern design problem formulated in \eqref{prob} has the following challenges:
(i) The constraint \eqref{cons-1} requires constant-modulus on all the elements of the combining vector, which is a non-convex constraint,
(ii) to respect the discrete phase shifter hardware constraint, $w_m$ can only take finite number of values based on all the possible phase shifts given by \eqref{cons-2},
(iii) the target UE's channel \textbf{$\mathbf{h}$ is assumed to be unknown}, since it is hard to acquire the CSI in practice, especially with analog/hybrid architectures, (iv) the channels of the interfering transmitters, i.e., \textbf{$\mathbf{h}_{k}, \forall k$, are also unknown}, since there is normally no coordination with  the  interfering transmitters, and  (v)  the possible hardware impairments are also assumed to be unknown.

Given these aforementioned difficulties, it is hard to solve \eqref{prob} using the conventional optimization methods \cite{Lorenz2005,Gesbert2010,Dahrouj2010}. An important observation, however, is that for a given combining beam $\bw$, evaluating the SINR requires only the power values (after combining) of the desired and interference signals, and does not require explicit knowledge about the channel vectors. Fortunately, it is less hard and more robust to acquire the receive power measurements for both the desired and interference signals, which requires much less training/signaling overhead and more relaxed synchronization constraints compared to the channel estimation process.
With this observation, we cast our problem as developing an online machine learning approach that learns how to design an interference-aware beam pattern $\bw$ that optimizes \eqref{prob}, \textbf{given only the receive power measurements} for the signal plus interference and noise, $\left|\bw^H \bh\right|^2P_x + \sum_{k=1}^K \left|\bw^H \bh_{k}\right|^2P_x + \sigma^2$, and the interference plus noise, $\sum_{k=1}^K \left|\bw^H \bh_{k}\right|^2P_x + \sigma^2$.

In the next section, we provide a detailed description of the proposed solution for tackling the beam learning optimization problem in \eqref{prob}.
Then, in \sref{sec:surrogate}, we  introduce a surrogate model assisted learning approach which achieves better sample efficiency as well as enhances  the deployment flexibility of the proposed interference-aware beam learning framework.

\section{Online Learning of Interference Aware Beam Pattern Design} \label{sec:BPL}

\begin{figure*}[t]
	\centering
	\includegraphics[width=.95\textwidth]{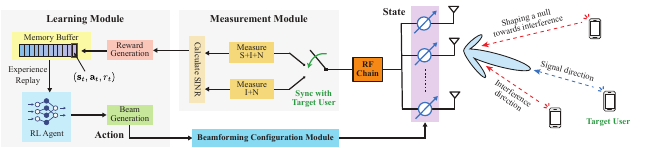}
    \caption{An illustration of the operation flow of the proposed interference-aware beam pattern learning solution, where the signal power is estimated by configuring the desired UE to transmit the signal in an on/off fashion.}
	\label{fig:on-off}
\end{figure*}

In this section, we present the proposed online reinforcement learning based interference-aware beam pattern learning approach.
The motivation of using reinforcement learning is two-fold:
First, the lack of the explicit channel knowledge renders most of the existing beamforming design approaches, such as \cite{Alkhateeb2016Frequency,Alrabeiah2020Neural}, infeasible.
Second, the beam design problem is essentially a search problem over a very huge space.
Hence, we consider leveraging the powerful exploration capability of reinforcement learning to efficiently navigate through this large space to find the optimal or near-optimal beam patterns.
Next, we first discuss the proposed system operation in \sref{subsec:Opt}.
Then, we provide the details of the proposed solution in \sref{subsec:BPL}.

\subsection{Practical System Operation} \label{subsec:Opt}

In this subsection, we discuss how to acquire the power measurements that are used for evaluating the objective function of the formulated beam design optimization problem \eqref{prob}, which will also be used in the proposed beam learning approach.
As will be discussed in \sref{subsec:BPL}, the proposed beam learning solution relies only on the power measurements in its operation. In particular, it needs to measure the power of the received signal from the target user as well as the interference power incurred from the other undesired transmitters.
Given that the BS can coordinate with its served UE to know when it is transmitting, this knowledge could be leveraged to enable the required power measurements.
To be more specific, to estimate the SINR performance of a certain beam $\tilde{{\bf w}}$, the BS first measures the interference plus noise level, i.e., $P_{\mathrm{I+N}}=\sum_{k=1}^K|\tilde{{\bf w}}^H{\bf h}_{k}|^2P_x + \sigma^2$, when the target UE is not transmitting.
Then, when the target UE starts transmitting reference signals, the BS uses the same beam to measure the signal plus interference plus noise level, i.e., $P_{\mathrm{S+I+N}}=|\tilde{{\bf w}}^H{\bf h}|^2P_x + \sum_{k=1}^K|\tilde{{\bf w}}^H{\bf h}_{k}|^2P_x + \sigma^2$.
The receive power of the target UE can hence be determined by subtracting the previously measured power $P_{\mathrm{I+N}}$ from the new power measurement $P_{\mathrm{S+I+N}}$, and the SINR can be approximately obtained as $(P_{\mathrm{S+I+N}}-P_{\mathrm{I+N}})/P_{\mathrm{I+N}}$.
To this end, it is worth mentioning that, in practice, zero power reference signals, such as the Sounding Reference Signal (SRS) that is not scheduled for any UE in the 5G NR, could be potentially leveraged to measure the interference plus noise level \cite{3gpp.38.802}, i.e., $P_{\mathrm{I+N}}$.
In the next subsection, we present the basic idea of the proposed online reinforcement learning approach for learning interference-aware beams based solely on these power measurements.

\subsection{Reinforcement Learning based Interference Aware Beam Pattern Design} \label{subsec:BPL}

Given the system operation in \sref{subsec:Opt}, we now describe our proposed reinforcement learning based solution for addressing the interference-aware beam pattern design problem \eqref{prob}.
To do that, we  first formulate our beam design problem as a reinforcement learning problem.
Then, we  present the proposed deep reinforcement learning architecture for solving this problem.

\noindent\textbf{Reinforcement Learning Formulation}:
To solve the problem with reinforcement learning, we first fit all the ingredients of problem \eqref{prob} into a general reinforcement learning framework as follows (as also illustrated in \fref{fig:on-off}):

\begin{itemize}
  \item \textbf{State:} We define the state ${\bf s}_t$ as a vector that consists of the phases of all the phase shifters at the $t$-th iteration, that is, ${\bf s}_t=\left[\theta_1, \theta_2, \dots, \theta_M\right]^T$. This phase vector can be converted to the actual combining vector ${\bf w}$ by applying \eqref{Analog}. Since all the phases in ${\bf s}_t$ are selected from $\boldsymbol\Psi$, and all the phase values in $\boldsymbol\Psi$ are within $(-\pi, \pi]$, \eqref{Analog} essentially defines a bijective mapping from the phase vector to the combining vector. In this paper, we will use the term ``combining phase vector'' to refer to this phase vector and use  the term ``combining  vector'' to refer to the actual combining vector.
  \item \textbf{Action:} We define the action ${\bf a}_t$ as the element-wise changes to all the phases in ${\bf s}_t$. Since the phases can only take values in $\boldsymbol\Psi$, a change of a phase represents the action that a phase shifter selects a value from $\boldsymbol\Psi$. Therefore, the action is directly specified as the next state, i.e., ${\bf a}_t = {\bf s}_{t+1}$, which can be viewed as a deterministic transition in the Markov Decision Process (MDP).
  \item \textbf{Reward:} We define a binary reward mechanism, i.e., the reward $r_t$ takes values from $\{+1, -1\}$. Since the objective of \eqref{prob} is to maximize the SINR performance, we compare the SINR achieved by the current combining vector, denoted as $\mathrm{SINR}_t$, with the previous one, i.e., $\mathrm{SINR}_{t-1}$. The reward is determined according to the following rule: $r_t=+1$, if $\mathrm{SINR}_t > \mathrm{SINR}_{t-1}$; $r_t=-1$, otherwise.
\end{itemize}
It is worth highlighting that the above reinforcement learning formulation is fully compatible with the original problem \eqref{prob} in the following aspects. First, since the state and action are directly specified as phase shifts of the discrete analog phase shifters, the constraints \eqref{cons-1} and \eqref{cons-2} are automatically satisfied. Second, to obtain the reward, the objective function of \eqref{prob}, i.e., the SINR performance, needs to be evaluated, which can be done in a way that does not rely on the channel state information of both the target user and the interfering transmitters, the details of which has been provided in \sref{subsec:Opt}.

\noindent\textbf{Deep Reinforcement Learning Architecture}:
Given the reinforcement learning formulation above for the interference-aware beam learning problem, we adopt an actor-critic based deep reinforcement learning architecture.
This follows the learning framework that we proposed earlier in \cite{Zhang2022Reinforcement}. In summary, both the actor and critic networks are implemented using elegant fully-connected (FC) feed-forward neural networks.
The input of the actor network is the state and the output is the action, while the critic network takes in the state-action pair and outputs the predicted Q value.\footnote{The detailed architectures and the parameters of the adopted neural networks are provided in \sref{sub:DLA}.}
Moreover, to respect the discrete phase shifter hardware constraint \eqref{cons-2}, we perform an element-wise quantization to make the predicted action a valid one. To be more specific, assume that $\widehat{{\bf a}}_t$ is the predicted action from the actor network at time $t$. Then, the action that finally gets implemented to the system is given by
\begin{equation}\label{Quant}
  [{\bf a}_t]_m = \mathop{\arg\min}_{\theta\in\boldsymbol{\Psi}}\left|[\widehat{{\bf a}}_t]_m - \theta\right|, ~ \forall m=1, \dots, M.
\end{equation}
%
It is worth emphasizing that such quantization operation is only activated when the system is actually implementing the predicted action by the actor network to obtain reward.
It is not involved in the training process of the actor network due to its non-differentiability.

Despite its full compatibility with the considered system as well as the 3GPP standards \cite{3gpp.38.802}, the proposed interference-aware beam learning solution still has two  drawbacks.
First, it  requires a relatively large number of iterations to find a qualified beam pattern, especially when the number of antennas is large. As a result, this incurs a large beam learning overhead, since these iterations are done over the air.
Second, as indicated by the objective function of \eqref{prob}, the SINR performance of a given beam is determined by two factors: (i) The desired beamforming gain and (ii) The effectiveness of suppressing the undesired interference.
However, the proposed solution does not fully leverage this information as it only focuses on the overall SINR performance.
It turns out that the decomposition of these two factors, as will be further discussed in the next section, makes the data sharing among the learning processes of different beams possible, which has the potential of improving the convergence behavior of the beam/codebook learning algorithm
With this motivation in mind, in the next section, we introduce a modified learning framework which includes a surrogate model that assists the beam learning process.
The adopted surrogate model better utilizes the underlying signal model and hence the sample efficiency (a measure for the number of real measurements) is further improved.
It also provides more deployment flexibility and enables other features such as data sharing.

\section{Surrogate Model Assisted Beam Learning Framework} \label{sec:surrogate}

In this section, we describe in detail the proposed surrogate model assisted interference-aware beam pattern learning framework.
The motivations of introducing the surrogate model are mainly two-folds.
First, it has the potential of improving the sample efficiency (i.e., reducing the number of interactions with the actual environment) of the learning process by leveraging the underlying signal models.
Second, it facilitates other more complex tasks (than learning a single beamforming vector), such as data sharing (which can be very useful in learning interference-aware beam codebook) and cooperative learning\footnote{For instance, as the system has full knowledge of its simulated environment, it can assign accurate reward to each agent. This has the potential of mitigating the non-stationary environment problem that exists in most of the multi-agent learning tasks.} (among multiple BSs to avoid interfering each other).
The overall objective is to have a \emph{simulated} environment that can provide the DRL agents with \emph{authentic} feedback as if the agents are interacting with the actual environment.
Next, we first introduce the adopted surrogate model in \sref{subsec:sur}.
Then, we provide more details about how to integrate the surrogate model with the RL beam learning algorithm in \sref{subsec:sur-ass}.
Finally, we discuss several practical aspects of the learning framework in \sref{subsec:prac}.

\subsection{Surrogate Model for Beam Pattern Learning} \label{subsec:sur}

In this subsection, we introduce the proposed surrogate model that assists the learning of interference-aware beams.
As can be seen from \sref{subsec:Opt}, in order to acquire the reward signal that is used for training the RL agent, the system needs to estimate two quantities, i.e., the signal power, $P_{\mathrm{S}} = \left|\bw^H \bh\right|^2P_x$, and the interference plus noise power, $P_{\mathrm{I+N}} = \sum_{k=1}^K \left|\bw^H \bh_{k}\right|^2P_x + \sigma^2$.
Therefore, correspondingly, there are two major components in the considered surrogate model that provide the agent with such information, i.e., an interference predictor and a signal predictor, as will be discussed in this subsection.

\subsubsection{The key idea of surrogate model}

The machine learning model that virtually interacts with the agent can be considered as a \textbf{surrogate model}.
This model is used to \emph{imitate} the behavior of the actual environment, aiming to reduce the expensive (sometimes, even impossible) actual evaluations of the design.
In this paper, we design the surrogate model with a particular emphasis on two important aspects:

\begin{itemize}

  \item \emph{Prediction accuracy:} As the name suggests, a surrogate model is essentially a prediction model which imitates (or predicts) the behavior (or response) of an unknown environment to a certain input action. Hence, being accurate is undoubtedly the most important property of the considered surrogate model.

  \item \emph{Data requirement:} Another important property of a surrogate model, in the considered interference-aware beam learning task, is data requirement. This refers to the amount of data that is required by the surrogate model for the training purposes, in order to reach a certain prediction accuracy constraint. Generally speaking, a surrogate model is more valuable if it requires less data to achieve a satisfactory performance.

\end{itemize}
With these important criterions in mind, we next describe the adopted surrogate model.
As mentioned before, the considered surrogate model consists of two major components, i.e., an interference prediction model and a signal prediction model.
Formally, the interference prediction model predicts the interference plus noise power based on the configuration of the receive combining vector, which can be expressed as
\begin{equation}\label{sur-int}
  \widehat{P}_{\mathrm{I+N}} = f_{\mathrm{in}}(\mathbf{w}; \boldsymbol{\Theta}_{\mathrm{in}}),
\end{equation}
where $\mathbf{w}\in\mathbb{C}^{M\times 1}$ is the input of the model, representing the designed receive combining vector, and the output is the predicted interference plus noise power, i.e., $\widehat{P}_{\mathrm{I+N}}\in\mathbb{R}$. The model is parameterized by $\boldsymbol{\Theta}_{\mathrm{in}}$.
Similarly, the signal prediction model predicts the signal power of a given receive combining vector, which can be written as
\begin{equation}\label{sur-sig}
  \widehat{P}_{\mathrm{S}} = f_{\mathrm{s}}(\mathbf{w}; \boldsymbol{\Theta}_{\mathrm{s}}),
\end{equation}
where $\widehat{P}_{\mathrm{S}}\in\mathbb{R}$ is the predicted signal power value and $\boldsymbol{\Theta}_{\mathrm{s}}$ denotes the model parameters.
It is worth mentioning that the architecture of $f_{\mathrm{in}}$ and $f_{\mathrm{s}}$ is not unique and is a design choice.
Next, we present two candidates that could be used in the considered beam learning task.

\subsubsection{Surrogate model architecture}

As mentioned before, the choice of $f_{\mathrm{in}}$ and $f_{\mathrm{s}}$ is not unique.
In this paper, we study two specific designs: (i) A model-based prediction architecture, and (ii) a fully-connected neural network based prediction architecture.

\noindent\textbf{Model-based architecture:}
The model-based architecture, as its name suggests, is inspired by the underlying \emph{signal model}.
For instance, as can be seen from the expression of the interference plus noise power, i.e., $P_{\mathrm{I+N}} = \sum_{k=1}^K \left|\bw^H \bh_{k}\right|^2P_x + \sigma^2$, it takes a quadratic form of the receive combining vector $\bw$.
To see this, by defining $\mathbf{H}=[\mathbf{h}_1, \mathbf{h}_2, \dots, \mathbf{h}_K]$, $P_{\mathrm{I+N}}$ can be expressed as
\begin{align}\label{in-pwr}
  P_{\mathrm{I+N}} & = \left\|\mathbf{H}^H\mathbf{w}\right\|_2^2P_x + \sigma^2, \\
    & = \mathbf{w}^H\left(P_x\mathbf{H}\mathbf{H}^H + \sigma^2\mathbf{I}\right)\mathbf{w}, \\
    & = \mathbf{w}^H\mathbf{A}\mathbf{w}, \label{in-pwr-quadratic}
\end{align}
where $\mathbf{A} = P_x\mathbf{H}\mathbf{H}^H + \sigma^2\mathbf{I}$.
The signal power can be expressed in the similar form, i.e., $P_{\mathrm{S}} = \mathbf{w}^HP_x\mathbf{h}\mathbf{h}^H\mathbf{w}$.
Therefore, the interference prediction network is essentially leveraged to learn the relationship \eqref{in-pwr-quadratic}.
Inspired by this, we design the interference prediction network with a focus on imitating the ``behavior'' of $\mathbf{A}$.
Specifically, the interference prediction network is chosen to take the following form
\begin{equation}\label{in-net}
  f_{\mathrm{in}}(\mathbf{w}) = \mathbf{w}^H\mathbf{Q}_{\mathrm{in}}\mathbf{Q}_{\mathrm{in}}^H\mathbf{w},
\end{equation}
where $\mathbf{Q}_{\mathrm{in}}\in\mathbb{C}^{M\times r_{\mathrm{in}}}$ with $r_{\mathrm{in}}$ being a hyperparameter.
Therefore, the parameter of the interference prediction network is essentially $\mathbf{Q}_{\mathrm{in}}$, i.e., $\boldsymbol{\Theta}_{\mathrm{in}} = \{\mathbf{Q}_{\mathrm{in}}\}$.
The signal prediction network takes the similar form, i.e., $f_{\mathrm{s}}(\mathbf{w}) = \mathbf{w}^H\mathbf{Q}_{\mathrm{s}}\mathbf{Q}_{\mathrm{s}}^H\mathbf{w}$, where $\mathbf{Q}_{\mathrm{s}}\in\mathbb{C}^{M\times r_{\mathrm{s}}}$ with $r_{\mathrm{s}}$ being a hyperparameter as well, which makes $\boldsymbol{\Theta}_{\mathrm{s}} = \{\mathbf{Q}_{\mathrm{s}}\}$.

\noindent\textbf{Fully-connected neural network based architecture:}
Despite being lightweight and a better fit to the signal model, the model-based architecture, fundamentally, suffers from any mismatch between the assumed signal model and the actual signal relationship.
For instance, there are normally unknown non-linearities in the practical hardware that undermine the validity of the assumed relationship between the receive combining vector and the interference plus noise power (similarly for the signal power).
As a result, the signal model cannot always be met and the model-based architecture will show up certain level of residual prediction errors that are very hard to be eliminated given the less powerful expressive capability of its architecture.
Motivated by this, we also investigate a more general architecture, which is built upon fully-connected neural network, given its powerful universal approximation capability \cite{NNUnivApprox}.
Specifically, both $f_{\mathrm{in}}$ and $f_{\mathrm{s}}$ are modeled with feed-forward fully-connected neural networks.
The detailed network parameters will be provided in \sref{sub:DLA}.

\subsubsection{Training dataset and loss function}

We denote the dataset used for training the interference prediction network as
\begin{equation}\label{in-data}
  \mathcal{D}_{\mathrm{in}} = \left\{\left(\mathbf{w}^{(n)}, P_{\mathrm{I+N}}^{(n)}\right)_{n=1}^{N_{\mathrm{in}}}\right\},
\end{equation}
where each data sample is comprised of a combining vector and its corresponding interference plus noise power value obtained from the actual environment, i.e., from the real measurement.
$N_{\mathrm{in}}$ is the total number of data samples in the dataset, i.e., $|\mathcal{D}_{\mathrm{in}}|=N_{\mathrm{in}}$.
And the dataset used for training the signal prediction network can be similarly denoted as
\begin{equation}\label{s-data}
  \mathcal{D}_{\mathrm{s}} = \left\{\left(\mathbf{w}^{(n)}, P_{\mathrm{S}}^{(n)}\right)_{n=1}^{N_{\mathrm{s}}}\right\},
\end{equation}
with $N_{\mathrm{s}}$ being its size.
We will discuss how to efficiently collect these datasets in \sref{subsec:prac}.

Since the target of these two networks is to predict the power values, we pose the learning problem as a regression problem conducted in a supervised fashion.
Furthermore, we employ mean squared error (MSE) as the training loss function.
Using the interference prediction network as an example, for the $n$-th data sample in $\mathcal{D}_{\mathrm{in}}$, the loss function is defined as
\begin{align}\label{loss}
  \mathcal{L}\left(P_{\mathrm{I+N}}^{(n)}, \widehat{P}_{\mathrm{I+N}}^{(n)}\right) & = \left|P_{\mathrm{I+N}}^{(n)} - \widehat{P}_{\mathrm{I+N}}^{(n)}\right|^2, \\
    & = \left|P_{\mathrm{I+N}}^{(n)} - f_{\mathrm{in}}(\mathbf{w}^{(n)}; \boldsymbol{\Theta}_{\mathrm{in}})\right|^2.
\end{align}
The loss function used for the signal prediction network is identical.

\begin{figure*}[t]
	\centering
	\includegraphics[width=.8\textwidth]{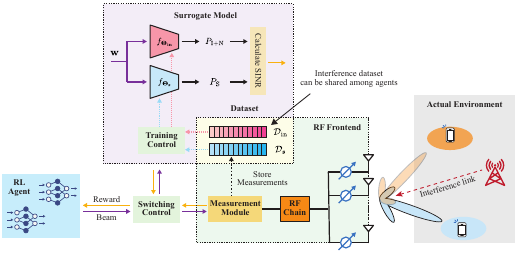}
    \caption{An illustration of the proposed surrogate model assisted interference-aware beam pattern learning framework.}
	\label{fig:surrogate-assisted}
\end{figure*}

\subsection{Surrogate Model Assisted Learning} \label{subsec:sur-ass}

In this subsection, we discuss how to integrate the surrogate model with the proposed RL based beam learning framework.
Since the surrogate model is essentially used to provide the RL agent with a simulated environment to interact with, it plays the same role as the actual environment.
However, in order to provide high quality \emph{synthetic} feedback, it requires training process that relies on the \emph{authentic} data collected from the actual environment.
Based on the trained surrogate model, the system can virtually evaluate its designed beams without measuring the physical signals.
Moreover, the system might require constantly switching between the surrogate model and the actual environment, triggered by the demand for the authentic data.
Next, we summarize the key components of the proposed surrogate model assisted beam learning.

\subsubsection{Initial interaction and data acquisition}

The system starts with the normal interaction between the RL agent and the actual environment.
To be more specific, upon forming a new beam $\tilde{{\bf w}}$, the BS follows the procedures presented in \sref{subsec:Opt} to estimate the interference plus noise power $P_{\mathrm{I+N}}$ and the signal power $P_{\mathrm{S}}$.
The reward signal used for RL agent learning will then be generated.
Moreover, these authentic power measurements together with the beam will be stored in the two datasets, i.e., $\mathcal{D}_{\mathrm{in}}$ and $\mathcal{D}_{\mathrm{s}}$, respectively.
During this interaction process, two initial datasets are established.

\subsubsection{Surrogate model training}

Based on the collected initial datasets $\mathcal{D}_{\mathrm{in}}$ and $\mathcal{D}_{\mathrm{s}}$, the two sub-networks of the surrogate model, i.e., the interference prediction network $f_{\mathrm{in}}$ and the signal prediction network $f_{\mathrm{s}}$, are trained in a supervised manner.
After the training process saturates, the surrogate model is ready to interact with the RL agent.

\subsubsection{Environment switching and virtual interaction}

The switching from the actual environment to the surrogate model is triggered based on multiple factors, which will be discussed in detail in \sref{subsec:prac}.
As a result, after the switching is finished, the reward signal required by the RL agent will be provided by the trained surrogate model instead of the actual environment.
The agent keeps interacting with the surrogate model until it does not improve, which marks the saturation of the agent learning and the end of the virtual interaction process.

\subsubsection{Demand based switching and active data acquisition}

The system might require executing the above steps multiple times, based on the achieved performance.
The motivation of such repetition can be summarized as follows.
From the model training perspective, the quality of the collected datasets, i.e., $\mathcal{D}_{\mathrm{in}}$ and $\mathcal{D}_{\mathrm{s}}$, has significant influence on the prediction accuracy of the trained surrogate model.
To be more specific, during the initial interaction process, most of the beams tried out by the agent are relatively random and hence have relatively poor quality in terms of SINR performance.
This means that the datasets are, intuitively speaking, biased towards the ``poor-quality'' beams.
As a result, the trained surrogate model will have relatively inaccurate predictions on the beams that actually have better performance.
The incurred residual prediction error will in turn influence the learning of the agent, leading to unsatisfactory performance.

However, as the policy of the RL agent gets improved over time, the actions performed by the agent, i.e., the beams, are more likely to be in the beam space where the achieved SINR is high.
Therefore, it is advisable to switching back to the actual environment to re-collect data (through agent-environment interaction).
Such active data acquisition can enhance the training datasets with ``high-quality'' beams.
Using those better data samples to refine the parameters of the surrogate model can help achieve higher prediction accuracy in the interested beam space, which further helps the learning of the agent.
By alternatingly performing these steps, the system has higher chance to collect data samples that are more useful for the agent learning, which has the potential of further enhancing both sample efficiency and learning convergence.
We show such interplay between the RL agent, actual environment and the surrogate model in \fref{fig:surrogate-assisted}.

\subsection{Practical Considerations} \label{subsec:prac}

In this subsection, we discuss some practical considerations of the proposed surrogate model assisted interference-aware beam pattern learning solution.

\subsubsection{Dataset collection}

An important observation is that for any given beam, its achieved signal power remains the same regardless of the presence of the interference signals.
This implies that the interference power dataset $\mathcal{D}_{\mathrm{in}}$ and the signal power dataset $\mathcal{D}_{\mathrm{s}}$, i.e., \eqref{in-data} and \eqref{s-data},
can be collected in a ``non-synchronized'' way, which facilitates the data collection in few cases.
For instance, on one hand, if the interference transmitters are not transmitting signals, the system can only collect data samples and store them into $\mathcal{D}_{\mathrm{s}}$. In this case, the interference plus noise power $P_{\mathrm{I+N}}$ becomes the noise power $P_{\mathrm{N}}$, and the signal power can be measured in a similar fashion.
On the other hand, if the interference happens aperiodically and sparse, the system can be dedicated to measure $P_{\mathrm{I+N}}$ whenever the interferers are present, i.e., without measuring the signal power during this period.

\subsubsection{Data sharing}

The aforementioned non-synchronized measurement strategy also implies that data sharing is possible.
To be more specific, if the interference transmitters are fixed, the collected interference dataset $\mathcal{D}_{\mathrm{in}}$ can be reused for different target UEs.
This has the potential of accelerating the learning process as well as reducing the memory requirement (i.e., to store the measurement dataset).
It is also particularly interesting when learning a codebook.

\subsubsection{Switching trigger}

Another important problem is how to design the conditions that control the switching between the surrogate model and the actual environment.
Although such criterion is a design choice and is normally determined by a variety of factors, it would be beneficial if the design can somehow reflect the intentions of introducing a surrogate model, which are:
(i) To support the continuous learning (when the actual environment cannot provide immediate feedback), and (ii) to reduce the expensive evaluations.
For instance, it is reasonable to switch to the actual environment when the training processes of both the surrogate model and the RL agent saturate.
However, when the surrogate model can provide very accurate predictions, the switching should be avoided to reduce the unnecessary overhead.

\subsubsection{Parallel computing}

It is also possible to perform the training of the surrogate model and the learning of the agent at the same time.
This can be done, for instance, by duplicating the surrogate model, where one of the duplicates (with its parameters being frozen) will be used to interact with the RL agent.
The other is trained with the up to date datasets, and after the training finishes, its parameters can be used to update the one that is interacting with the RL agent.
This implies that the second and third steps mentioned in the previous subsection can be executed in parallel, which can help improve the convergence of the proposed solution.

\subsubsection{Quantized measurements}

The proposed surrogate model also supports the cases when the measurements are quantized.
In such case, the output layer of both interference prediction network and signal prediction network can be modified to be a classification layer.

\section{Simulation Results} \label{sec:Simu}

In this section, we numerically evaluate the performance of the proposed reinforcement learning based interference-aware beam pattern design approach. We will first describe the adopted simulation setup in \sref{subsub:SSet}. Then, in \sref{sub:DLA}, we provide more details about the adopted architectures of the deep learning models as well as the training procedures. Finally, in \sref{subsub:NResult}, we present the numerical results of the proposed solutions.

\subsection{Simulation Setup} \label{subsub:SSet}

In this simulation, we consider the case where a BS receiver adopts uniform linear array (ULA) with half-wavelength antenna spacing.
Each antenna of the ULA is followed by a 3-bit analog phase shifter.
Besides, for a better demonstration, we adopt the following simulation steps:
(i) We generate the channel of the target user based on \eqref{ch}, where, for simplicity, we consider the case when the user only has a line-of-sight (LOS) connection with the BS, i.e., $L=1$ in \eqref{ch};
(ii) We then learn a beam pattern assuming there is no interference and this learned beam is referred to as \textbf{``interference-unaware'' beam}, since it solely focuses on maximizing the combining gain of the desired signal;
(iii) After this beam is learned, we \emph{intentionally} position the interfering transmitters at the directions aligning with the strongest side lobes of the learned beam and also assume that they only have LOS channels with the considered BS, which causes non-negligible interference;
and (iv) We finally take the interference into account and re-design an \textbf{``interference-aware'' beam} that learns how to manage the interference in such a way that improves the SINR performance.

\subsection{Deep Learning Models and Training Procedures} \label{sub:DLA}

In this subsection, we provide more details about the adopted deep learning architectures for both the DRL agent and the surrogate model. We also provide the parameters regarding the training processes of these models.

\subsubsection{DRL agent architecture}

Since the input of the actor network is the state and the output is the action, the size of both the input and output of the actor network is $M$, i.e., the number of antennas.
The critic network takes in the state-action pair and outputs the predicted Q value and hence it has an input size of $2M$ and an output size of $1$.
Both the actor and critic networks have two hidden layers in our proposed architecture, with the size of the first hidden layer being $16$ times of the input size and the size of the second hidden layer being $16$ times of the output size in both networks.
All the hidden layers are followed by the batch normalization layer for an efficient training experience and the Rectified Linear Unit (ReLU) activation layer.
The output layer of the actor network is followed by a Tanh activation layer scaled by $\pi$ to make sure that the predicted phases are within $(-\pi, \pi]$ interval.
The output layer of the critic network is a linear layer.
Moreover, it is worth mentioning that we adopt the same DRL architecture for both solutions, regardless of having surrogate model or not.

\subsubsection{Surrogate model architecture}

We describe the two different architectures of the surrogate model studied in this paper.
Also, as the signal prediction network and the interference prediction network have identical architecture in both solutions (i.e., model-based solution and fully-connected neural network based solution), for brevity, we only use the interference prediction network as an example.

\noindent\textbf{Signal model-based prediction network}
As mentioned before in \eqref{in-net}, the interference prediction network is essentially devised to take a quadratic form of the combining vector determined by a positive semi-definite matrix $\mathbf{Q}_{\mathrm{in}}\mathbf{Q}_{\mathrm{in}}^H$, leaving the matrix $\mathbf{Q}_{\mathrm{in}}$ to be the model parameter.
Moreover, $\mathbf{Q}_{\mathrm{in}}$ has a shape of $M\times r_{\mathrm{in}}$ with $M$ being the number of antennas and $r_{\mathrm{in}}$ being a hyper-parameter.
The choice of $r_{\mathrm{in}}$ is empirically guided by the following rules: (i) $r_{\mathrm{in}}$ should not be too large as it will increase the model complexity and hence the required amount of training data; (ii) $r_{\mathrm{in}}$ should not be too small as it will limit the expressive capability of the model, leading to unsatisfactory prediction accuracy.

\noindent\textbf{Fully-connected neural network based prediction network:}
We adopt the fully-connected neural network with two hidden layers to be the interference prediction network.
The input layer of the network has $M$ neurons, which is equal to the number of antennas.
The output layer of the network has only one neuron with linear activation.
Both hidden layers have $M^\prime$ neurons.
Similar to $r_{\mathrm{in}}$ in the model-based architecture, the selection of $M^\prime$ needs to strike a balance between model complexity and model expressive capability.
Moreover, all the hidden layers are followed by the batch normalization layer and ReLU activation layer.

\begin{table}[t]
\caption{Hyper-parameters for surrogate model training}
\centering
\begin{tabular}{c|c|c}
  \hline
  \hline
  \textbf{Parameter} & \textbf{Model-based} & \textbf{FC-based} \\
  \hline
  Batch size & 512 & 512 \\
  Number of epochs & 500 & 500 \\
  Optimizer & Adam & Adam \\
  Initial learning rate & $1\times10^{-1}$ & $1\times10^{-2}$ \\
  Learning rate schedule & $0.1$@$\left[50, 300, 400\right]$ & $0.1$@$\left[100, 300, 400\right]$ \\
  \hline
  \hline
\end{tabular}
\label{TrParam}
\end{table}

\subsubsection{Training parameters}

As mentioned before, the surrogate model is trained in a supervised fashion, based on the collected power datasets, i.e., $\mathcal{D}_{\mathrm{in}}$ and $\mathcal{D}_{\mathrm{s}}$.
Moreover, the interference prediction network and the signal prediction network are independently trained.
However, for the same type of surrogate model, i.e., either model-based or fully-connected neural network based, we adopt the same training parameters for interference and signal prediction networks.
We summarize the detailed hyper-parameters used for training the surrogate models in \tref{TrParam}.


\subsection{Numerical Results} \label{subsub:NResult}

In this subsection, we provide the simulation results of the proposed interference-aware beam learning solutions.
Moreover, to better present the results, in \sref{subsubsec:simu-1}, we first evaluate the reinforcement learning based beam design solution proposed in \sref{sec:BPL} that keeps interacting with the actual environment. This is to demonstrate the achieved performance by the proposed beam learning algorithm without knowing the channel knowledge.
Then, in \sref{subsubsec:simu-2}, we test the surrogate model assisted beam design framework proposed in \sref{sec:surrogate}, with a focus on evaluating the validity and efficacy of using surrogate model to reduce the beam learning overhead, as well as comparing different surrogate model architectures.

\subsubsection{Interference nulling without knowing the channels} \label{subsubsec:simu-1}

\begin{figure*}[t]
	\centering
    \subfigure[Interference-unaware]{ \includegraphics[width=0.305\textwidth]{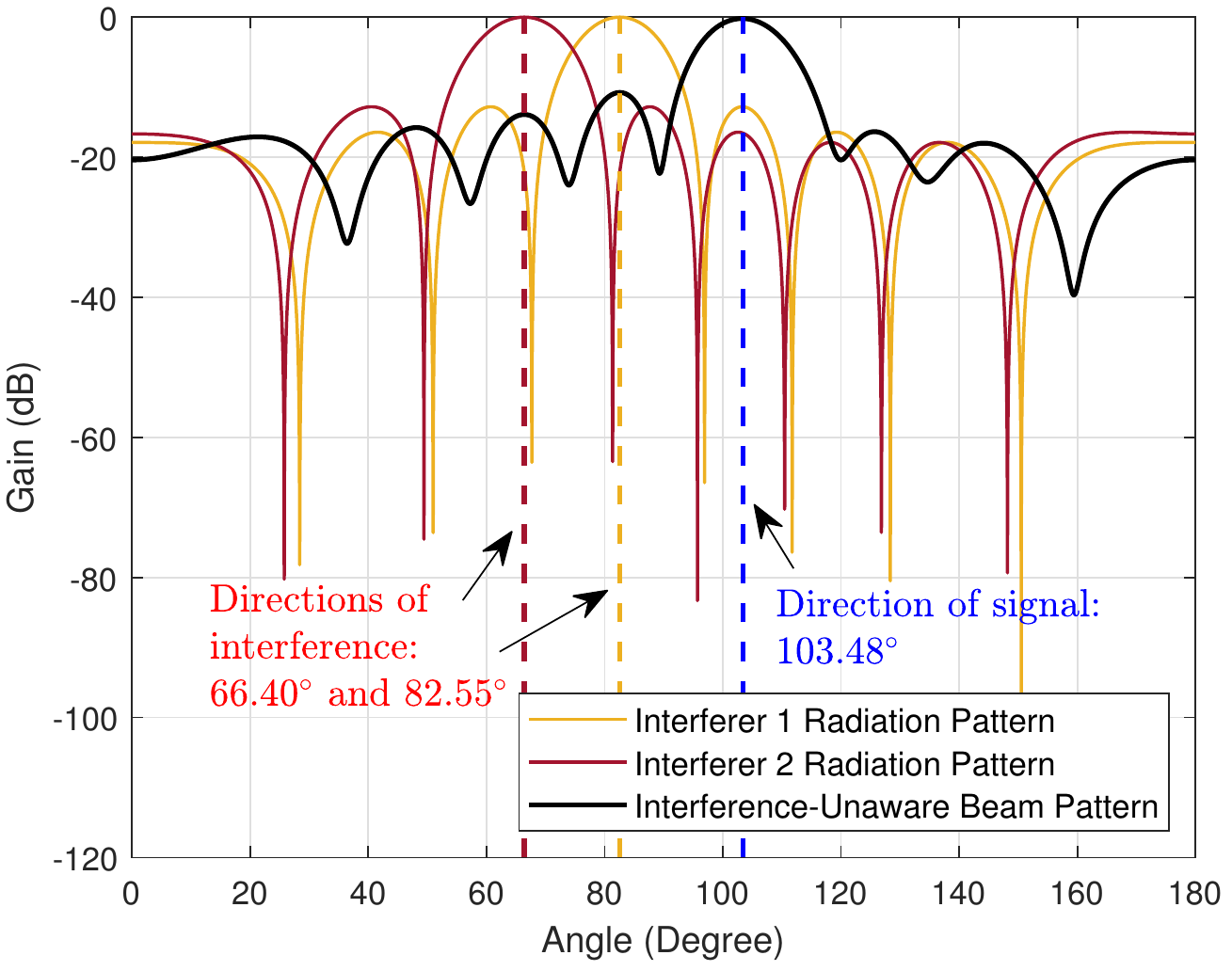} \label{two-interf-fig:a} }
	\subfigure[Interference-aware]{ \includegraphics[width=0.305\textwidth]{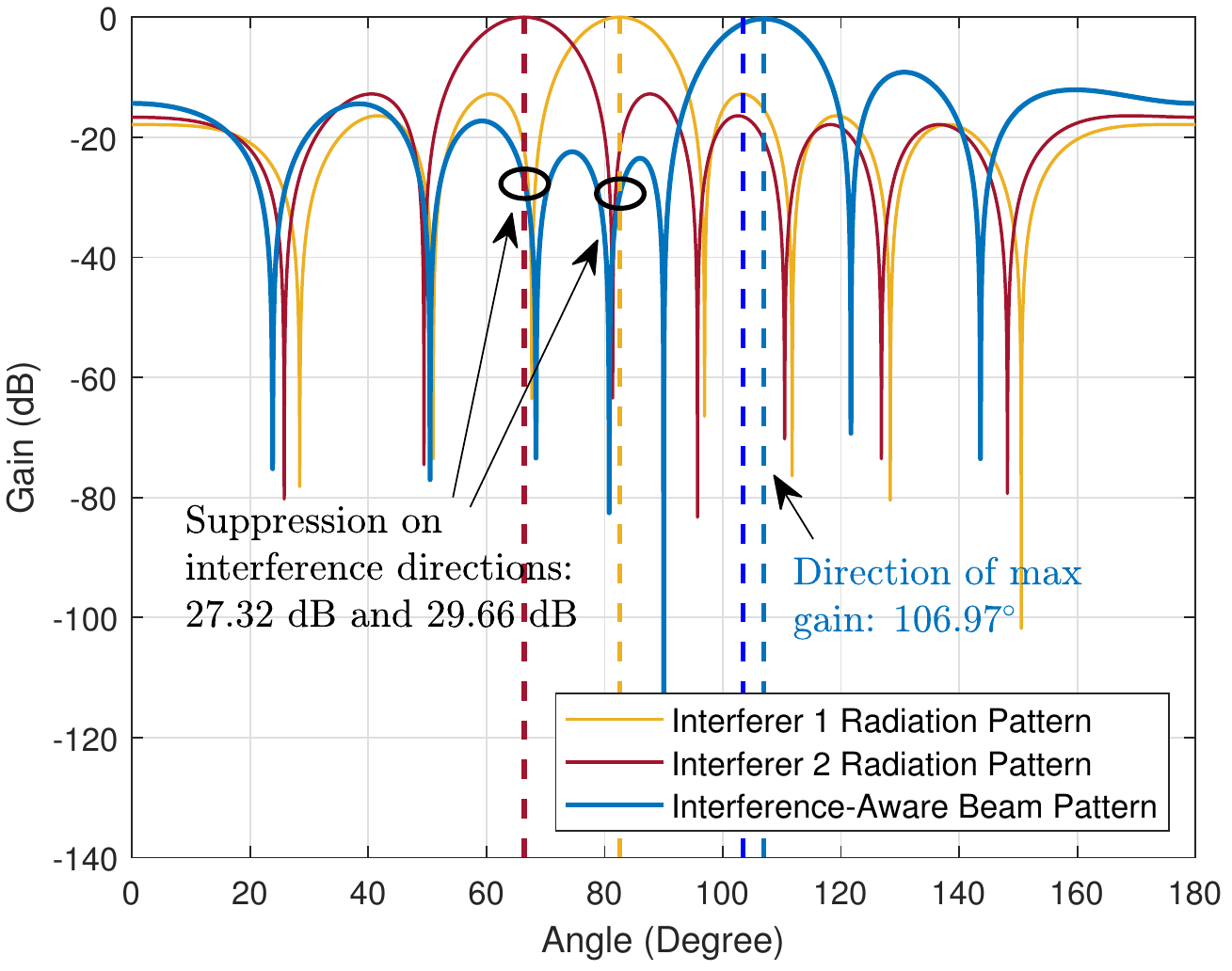} \label{two-interf-fig:b} }
    \subfigure[Learning process]{ \includegraphics[width=0.32\textwidth]{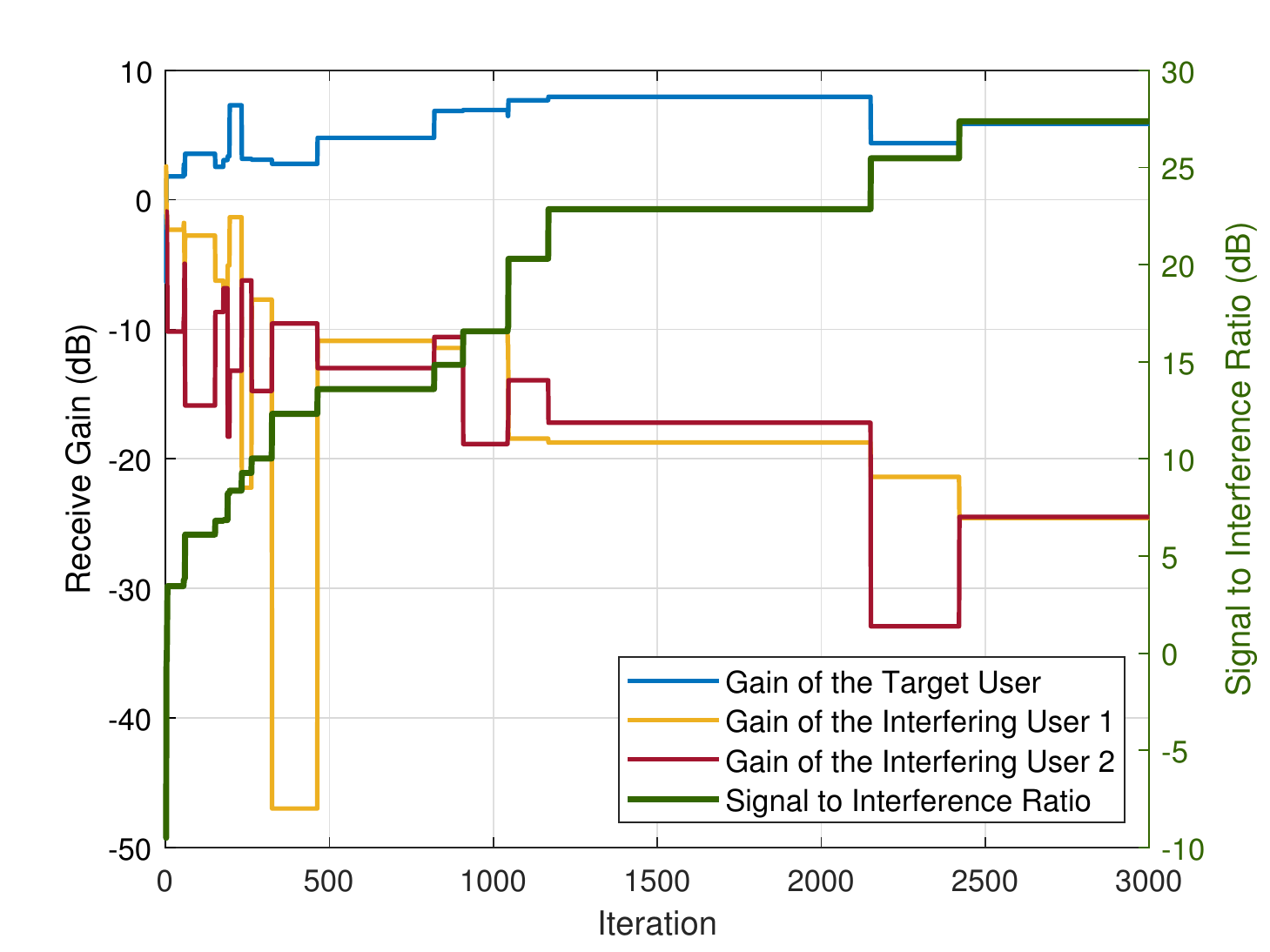} \label{pf-vs-iter} }
	\caption{The beam pattern learning results in an environment with two interfering transmitters, where (a) shows the learned beam pattern when ignoring the surrounding interfering transmitters, and (b) shows the interference-aware beam pattern. (c) shows the interference-aware beam pattern learning process.}
	\label{two-interf}
\end{figure*}

Based on the aforementioned simulation setup and deep learning architecture, in \fref{two-interf}, we demonstrate the learning results when there are two interfering transmitters.
We show the beam patterns learned with and without taking the interference into account, together with the receive patterns (i.e., the distribution of receive power strength in angular domain from the BS's perspective) of the selected interfering sources.
As shown in \fref{two-interf-fig:a}, the two interferers are present at the directions aligning with the two most strongest side-lobes of the interference-unaware beam, which incurs significant interference and causes performance degradation.
The learned interference-aware beam is plotted in \fref{two-interf-fig:b}. As can be seen, \textbf{the interference-aware beam shapes nulls that have very low receive gains at the directions of  the interferers, which nearly eliminates the severe interference.} To be more specific, in the interference-unaware case, the signal-to-interference ratio (SIR) levels are $10.56$ dB and $13.71$ dB with respect to the two interfering transmitters. By contrast, the SIR levels are improved to $28.63$ dB and $26.28$ dB when using the interference-aware beam, which only incurs a loss of $0.8348$ dB for the combining gain of the target user.

In \fref{pf-vs-iter}, we show how the combining gains of the received signals from the target user and the interfering transmitters are changing as the learning proceeds, as well as the overall SIR performance. As can be seen, the combining gain of the target user and the combining gains of the two interfering transmitters start from almost the same level, since a random beam is used as the starting point. As learning proceeds, the combining gain of the target user maintains, generally speaking, an increasing trend, while the combining gains of the two interfering transmitters are gradually decreasing. The overall SIR, however, maintains a monotonically increasing trend.
Furthermore, as can be observed from the figure, the combining gain of the target user has a high spike (achieved by a certain learned beam) at the beginning of the learning process. However, despite the good performance on the target user, that beam also incurs strong interference from other undesired transmitters, hence resulting in an unsatisfactory SIR performance. Therefore, it is finally replaced by other beams that have slightly lower combining gain to the target user but are very effective in suppressing the interference.
\fref{pf-vs-iter} also shows that \textbf{with only around $1000$ iterations, the SIR performance improved from around $-10$ dB to around $20$ dB}, without knowing the channels (for both the target user and the interfering transmitters).

\begin{figure*}[t]
	\centering
    \subfigure[Signal and interference prediction ($M=8$)]{ \includegraphics[width=0.45\textwidth]{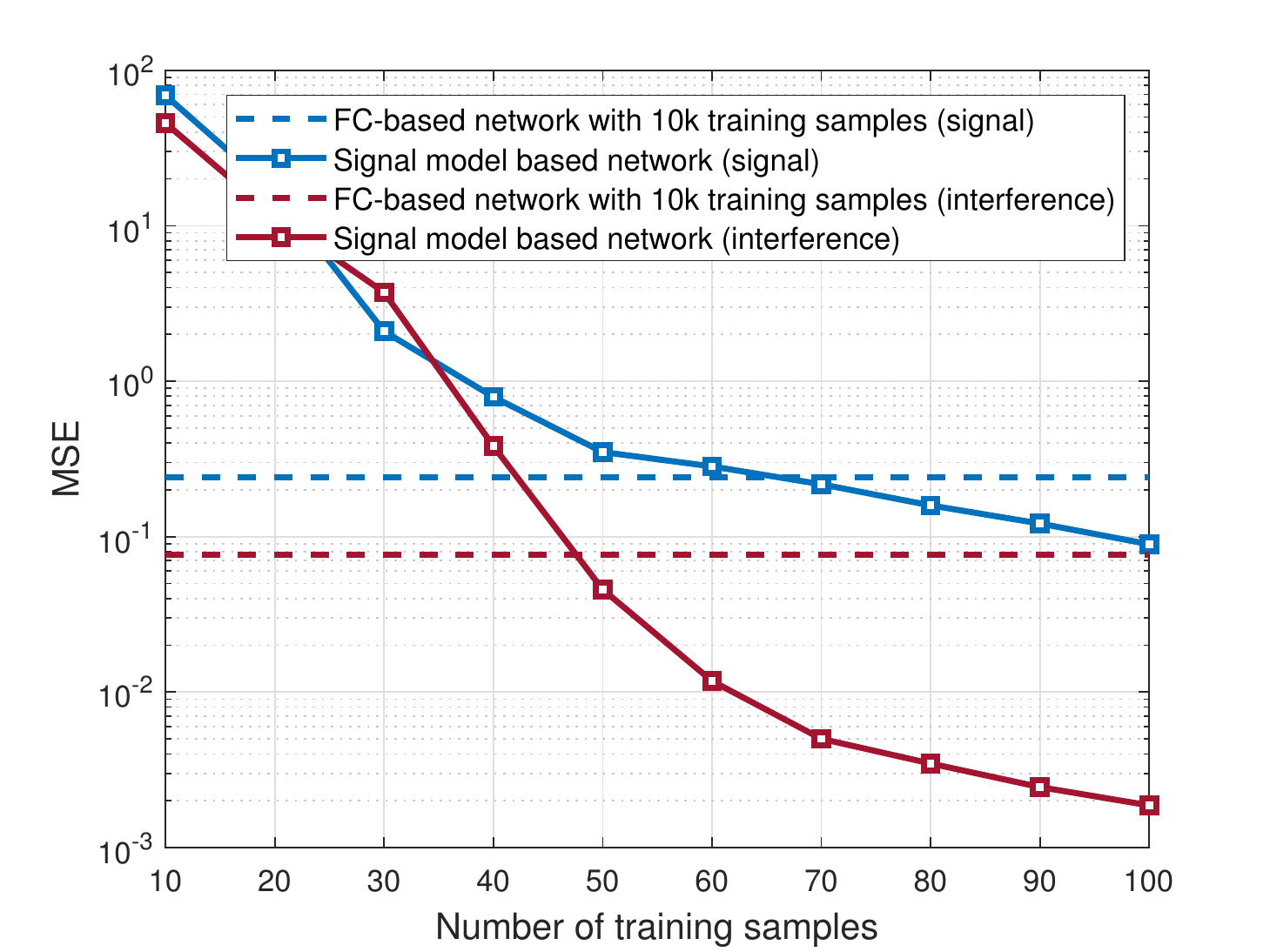} \label{acc-vs-sample-8} }
    \subfigure[Signal and interference prediction ($M=256$)]{ \includegraphics[width=0.45\textwidth]{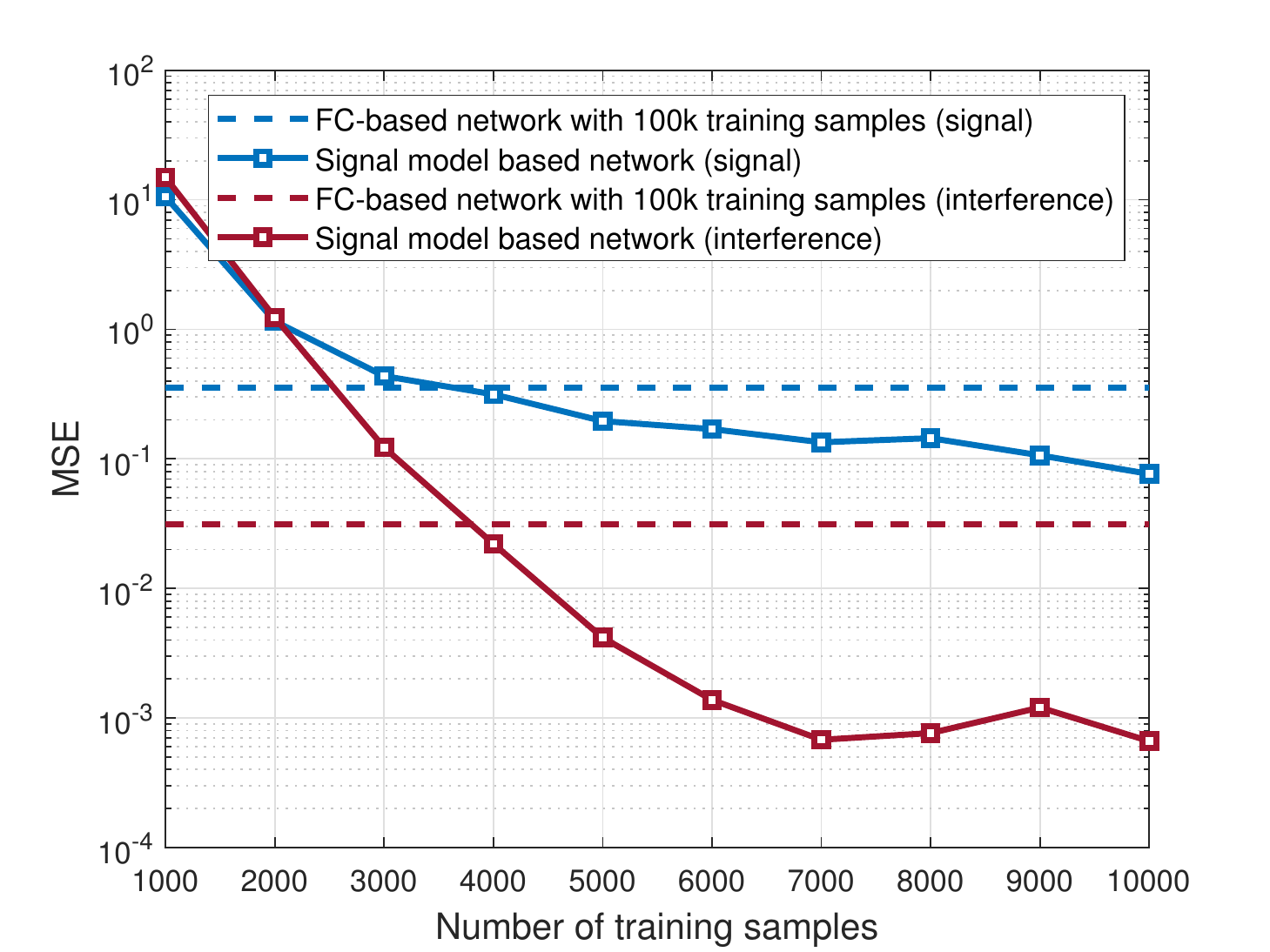} \label{acc-vs-sample-256} }
	\caption{The prediction accuracy of different surrogate model architectures. It shows that the proposed signal model-based prediction network requires much less data samples to outperform the FC-based prediction network in both cases, i.e., the base station equipping (a) $M=8$ antennas, and (b) $M=256$ antennas.}
	\label{acc-vs-sample}
\end{figure*}

\begin{figure*}[t]
	\centering
    \subfigure[Interaction with the actual environment]{ \includegraphics[width=0.45\textwidth]{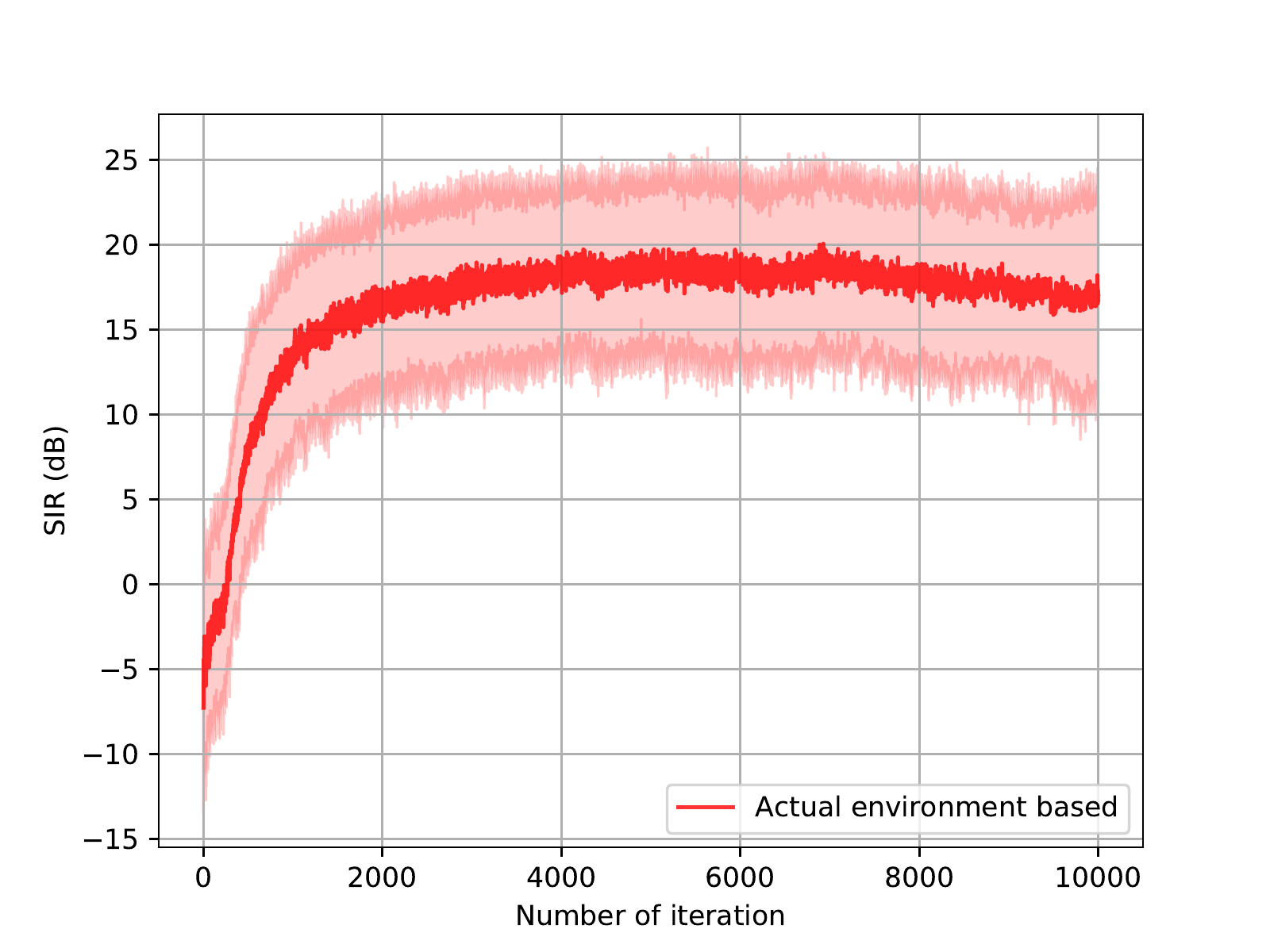} \label{actual-based} }
    \subfigure[Interaction with the surrogate model]{ \includegraphics[width=0.45\textwidth]{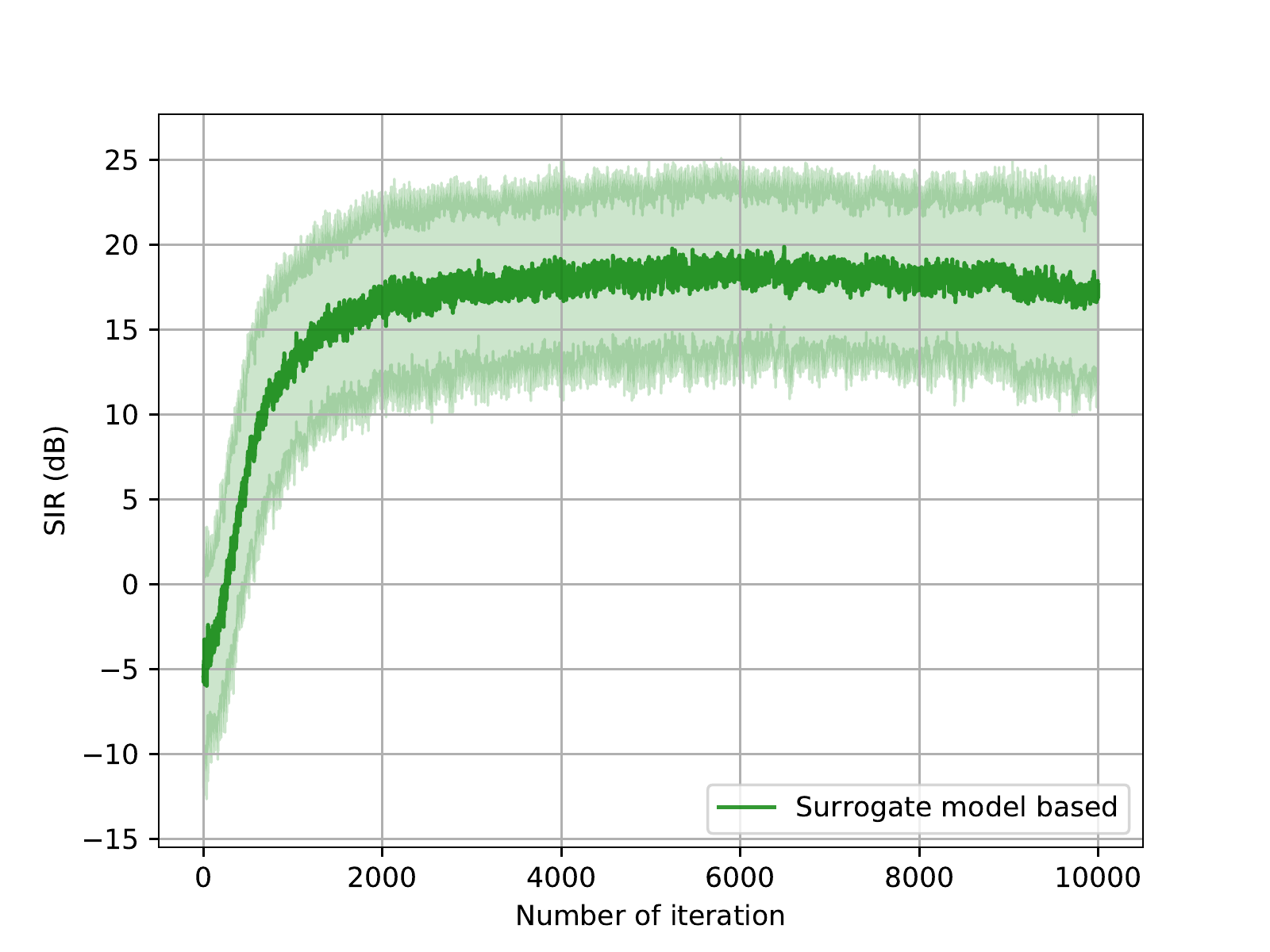} \label{surrogate-based} }
	\caption{The learning experience of the DRL agent when interacting with (a) the actual environment and (b) the surrogate model trained with $1000$ data samples.}
	\label{fig:iteract}
\end{figure*}

\subsubsection{Surrogate model assisted learning} \label{subsubsec:simu-2}

We also evaluate the performance of the surrogate model assisted learning framework, which has the potential of significantly reducing the number of interactions with the actual environment.

In \fref{acc-vs-sample}, we first evaluate the prediction accuracy of the two proposed prediction network architectures, which provides insight on how much data samples are required in order to have a reasonable performance as well as the practicality of the solutions.

We show the prediction accuracy of both the signal power and the interference power.
As can be seen, the signal model-based architecture requires much less data samples to achieve higher prediction accuracy than the FC-based architecture trained with much more data samples.
For instance, as indicated in \fref{acc-vs-sample-8}, with only $50$ samples, the signal model-based prediction architecture can achieve even more accurate interference prediction than the FC-based architecture trained with $10,000$ samples.
\textbf{This saves almost $99.5\%$ of the measurements, yielding a more sample-efficient solution for the practical system deployment.}
Moreover, as there are more data samples, the prediction accuracy of the signal model-based architecture also gets improved quite significantly.
Such performance is achieved by better leveraging the underlying signal relationships and hence the model parameters are essentially searched over a much smaller space.
Finally, by comparing \fref{acc-vs-sample-8} and \fref{acc-vs-sample-256}, it can be observed that with more antennas, the system needs to collect more data samples to train the prediction networks in order to maintain the similar prediction accuracy when the number of antennas is small.

The trained surrogate model is utilized to interact with the DRL agent, aiming to reduce the expensive actual measurements conducted by the hardware.
In \fref{fig:iteract}, we show the performance of the DRL agent when interacting with the actual environment as well as the surrogate model.
The training of the DRL agent is repeated for $100$ times and the average performance as well as the standard deviation are reported in \fref{fig:iteract}.
We test the performance of a system with $8$ antennas, and the surrogate model is trained using $1,000$ data samples, i.e., $|\mathcal{D}_{\mathrm{in}}|=|\mathcal{D}_{\mathrm{s}}|=1000$.
As can be seen, the learning experience based on the surrogate model is quite similar to that of the one based on the actual environment.
This empirically shows the effectiveness of using the surrogate model in training the DRL agent.
As a result, although the DRL agent requires almost a total number of $5,000$ interactions with the environment to converge, \textbf{in the surrogate model assisted learning framework, all these interactions are with the surrogate model and hence the expensive evaluations on the real hardware are avoided.}

\section{Real Measurement Results} \label{sec:Real}

In this section, we further evaluate the performance of the proposed  interference-aware beam pattern learning algorithm in \sref{subsec:BPL} using a real-world mmWave prototyping platform.

\subsection{Hardware Platform Description}

\begin{figure*}[t]
	\centering
	\subfigure[Prototyping setup in an outdoor environment]{ \includegraphics[width=.46\textwidth]{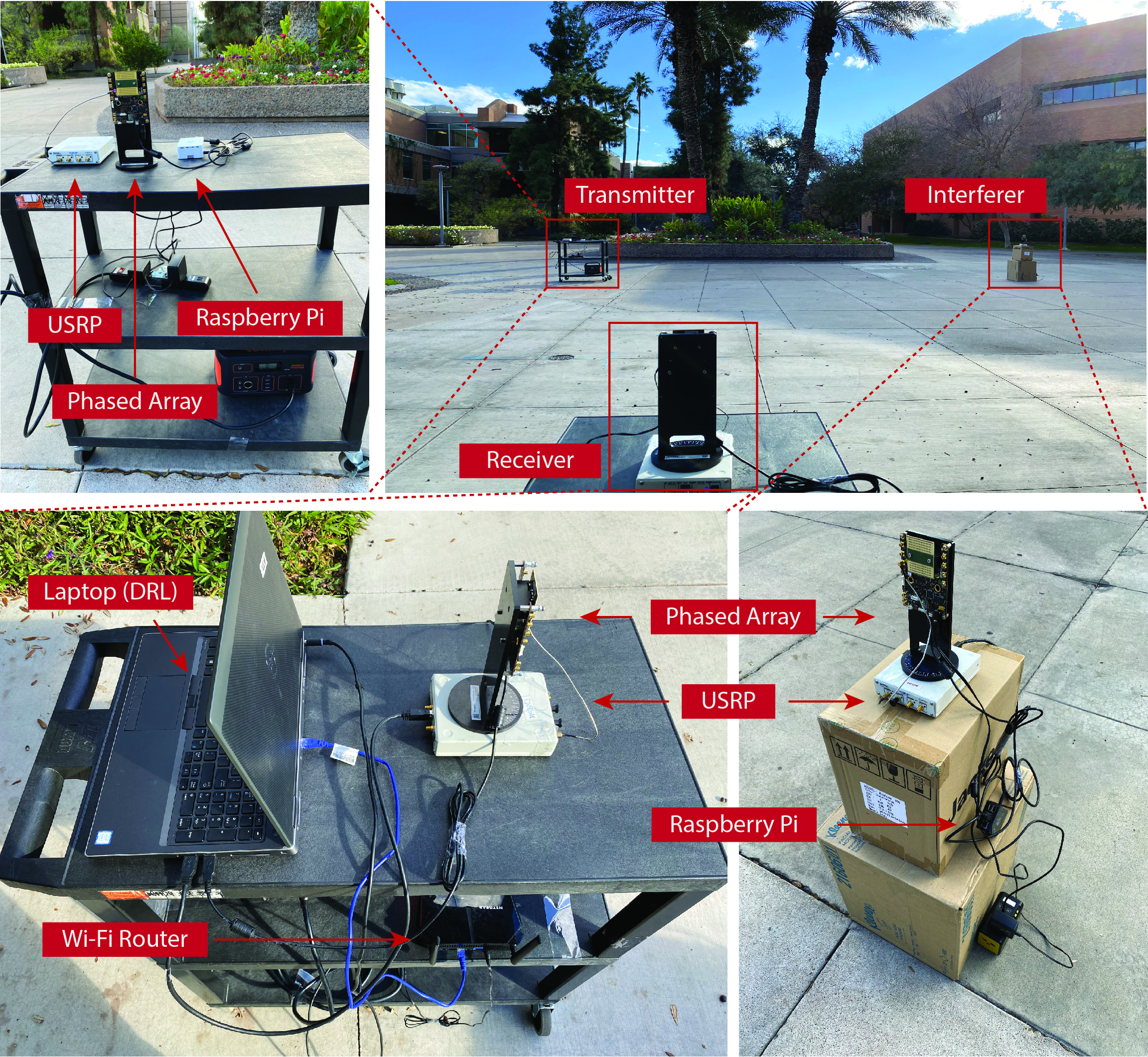} \label{setups} }\hspace{5pt}
    \subfigure[Measurement campaign]{ \includegraphics[width=.42\textwidth]{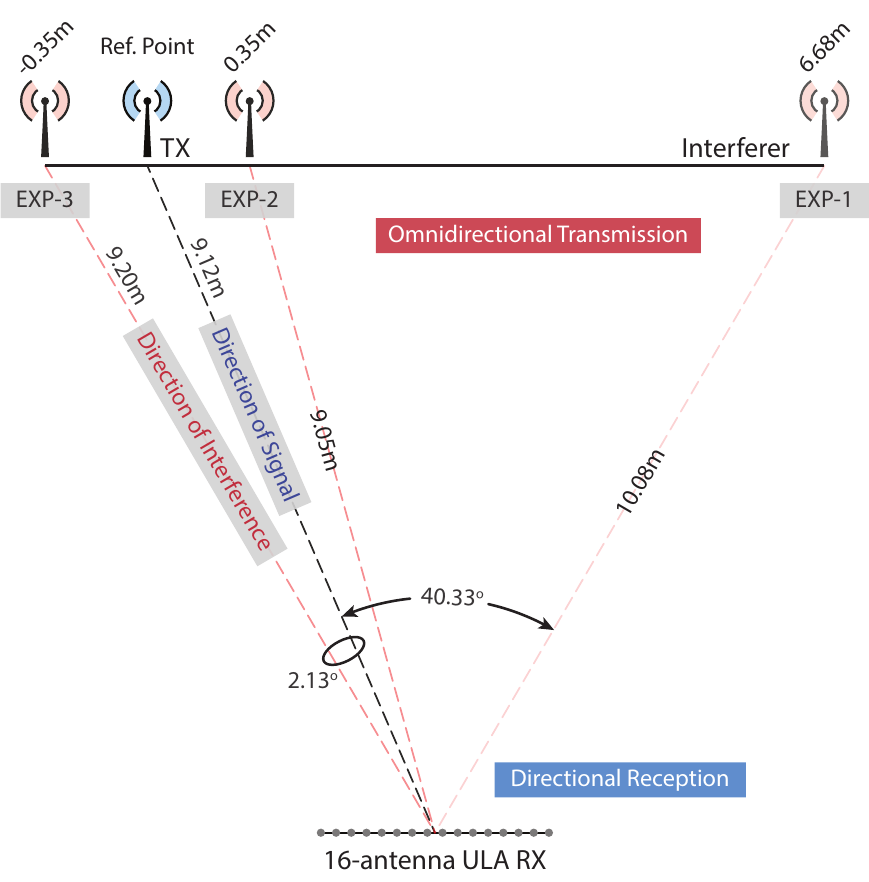} \label{outdoor-exp} }
	\caption{The prototyping setup and the outdoor measurement environment for evaluating the proposed interference-aware beam pattern design algorithm. The adopted setup consists of a receiver, a desired transmitter and an interferer, as shown in (a). The upper right figure in (a) shows the EXP-1 of the conducted measurement campaign, as depicted in (b), where we provide an illustration of the relative positions of the receiver, transmitter and interferer in the outdoor measurements.}
	\label{proto-setup}
\end{figure*}

As shown in \fref{setups}, we build a test platform comprised of a receiver, a transmitter, and an interferer.
The radio frontend of all three components is the same type of mmWave phased array, which employs a $16$-antenna uniform linear array (ULA) and transmits/receives signals at an operating frequency of $62.64$ GHz.
The control units of the transmitter and the interferer are identical, while the control unit of the receiver includes a laptop.
The laptop is used for several tasks: (i) It controls the phased array at the receiver; (ii) It executes the deep reinforcement learning algorithm; (iii) It connects to a wireless router and can remotely control the transmitter and the interferer. During the measurement, it controls the on/off status of the transmitter.
It is worth mentioning that although the transmitter and interferer are equipped with phased arrays, they both transmit signals in an \textbf{omnidirectional} way for an effective and fair evaluation of the proposed algorithm.
For the phased array at the receiver, only $2$ bits are used for the phase encoding of each phase shifter to form the \textbf{directional} beam, which means that the signal received by each antenna can only be adjusted with $4$ different phase values.

\subsection{Experiment Description}

In this subsection, we describe in detail how the experiments are designed to effectively evaluate the performance of the proposed algorithm.
As can be seen in \fref{setups}, we consider the scenario where both transmitter and interferer have a direct LOS connection with the receiver.
To better reflect the interference suppression capability, we first turn on only the transmitter and use the algorithm proposed in \cite{Zhang2022Reinforcement} to learn the ``interference-unaware'' beam, which focuses only on maximizing the receive power from the desired transmitter.
Then, we turn on the interferer and run the algorithm proposed in \sref{sec:BPL}. This forms the ``interference-aware'' beam, which focuses on maximizing the SINR performance.
For a better understanding of the measurement results, we also depict the relative positions of the receiver, transmitter and interferer in \fref{outdoor-exp}. It is worth pointing out that, during the measurement, we fixed the positions of the transmitter and receiver, which maintains a distance of $9.12$m. We change the position of the interferer to get multiple angle differences that are of interest between the two LOS links. For example, as shown in \fref{outdoor-exp}, in the EXP-1, the distance between the receiver and the interferer is $10.08$m, and the distance between the transmitter and the interferer is $6.68$m, which forms a angle difference of around $40.33^\circ$ from the receiver's point of view.
Finally, we also visualize all the learned beams by measuring their patterns in an anechoic chamber as shown in \fref{exp-unaware:c}, which provides useful information in understanding and validating the achieved performance.

\subsection{Measurement Results}

\begin{figure*}[t]
	\centering
    \subfigure[Power Measurement]{ \includegraphics[width=0.29\textwidth]{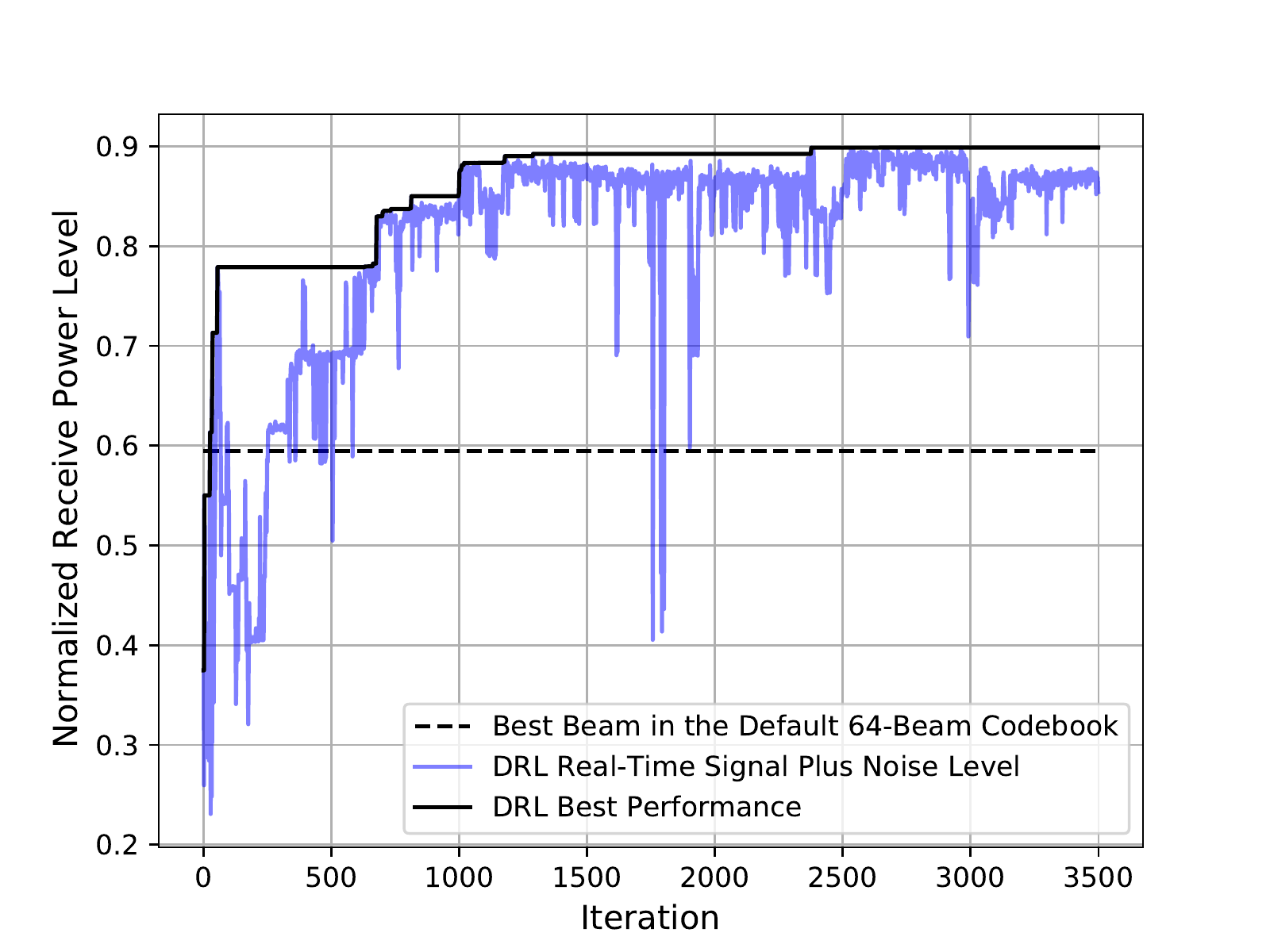} \label{exp-unaware:a} }
    \subfigure[Anechoic Chamber Setup]{ \includegraphics[width=0.42\textwidth]{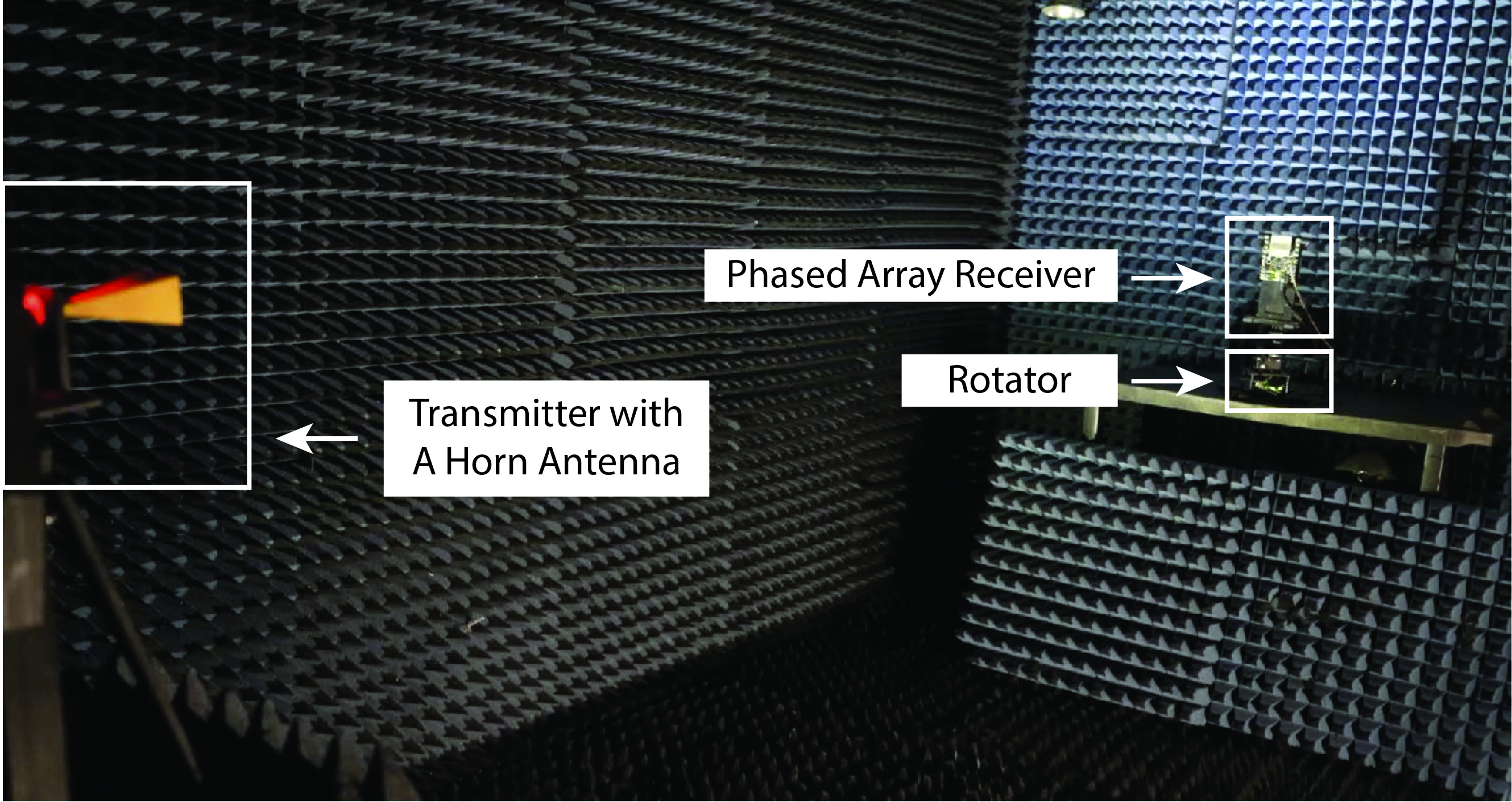} \label{exp-unaware:c} } 
    \subfigure[Beam Pattern]{ \includegraphics[width=0.22\textwidth]{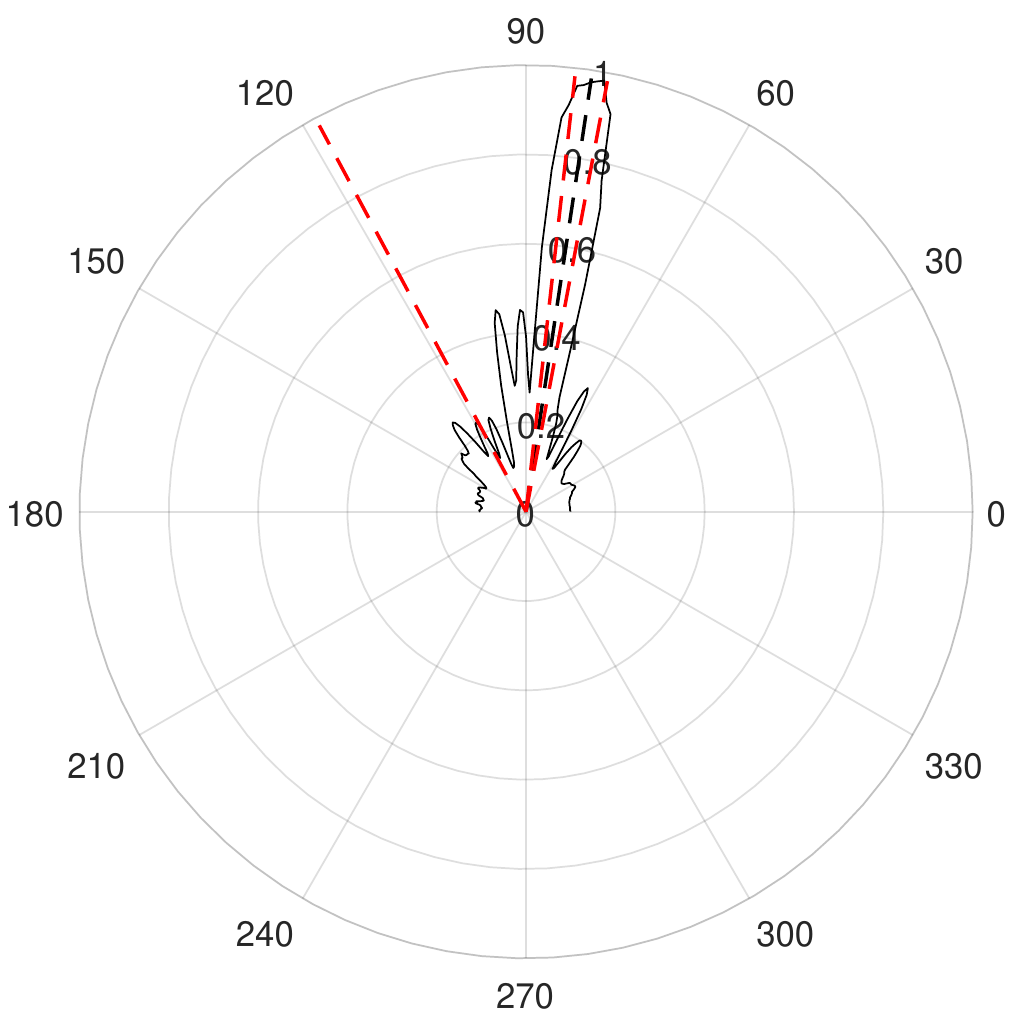} \label{exp-unaware:b} }
	\caption{The learning results of the interference-unaware beam pattern, where (a) shows the real-time power measurement, (b) shows the anechoic chamber setup for measuring the learning beam pattern, and (c) shows the learned beam pattern with the black dashed line representing the direction of the desired signal and the red dashed lines representing the directions of the interfering sources which will be presented later.}
	\label{exp-unaware}
\end{figure*}

\begin{table}[t]
\caption{Performance of the interference-unaware beam}
\centering
\begin{tabular}{c|ccc}
  \hline
  \hline
  & EXP-1 & EXP-2 & EXP-3 \\
  \hline
  \textbf{Angular Separation} & $40.33^\circ$ & $2.16^\circ$ & $2.13^\circ$ \\
  \textbf{Signal Plus Noise Level} & $0.92$ & $0.86$ & $0.903$ \\
  \textbf{Interference Plus Noise Level} & $0.32$ & $0.915$ & $0.94$ \\
  \textbf{SIR Performance} & $6.96$ dB & $-0.33$ dB & $-0.22$ dB \\
  \textbf{INR Performance} & $-0.47$ dB & $6.56$ dB & $6.29$ dB \\
  \hline
  \hline
\end{tabular}
\label{Measure-Table}
\end{table}

In \fref{exp-unaware:a}, we plot the learning process of the interference-unaware beam pattern, where the real-time performance of the DRL-based beam pattern learning algorithm is presented. To better understand the learning process and make sure that the learning result is meaningful, we compare the performance achieved by the learned beam with a built-in beam codebook. To be more specific, the phased array adopted in the experiment includes a default codebook that has $64$ beams. This codebook is essentially a beamsteering-like codebook with its beams covering $-45^\circ$ to $+45^\circ$ azimuth angular space. In order to find the beam with the best performance, we perform a beam sweeping and calculate the receive power after combining by each of the beams. The one that achieves the highest receive power is determined as the best beam. As can be seen in the \fref{exp-unaware:a}, the learned beam finally achieves a normalized receiver power of around $0.9$, significantly outperforming the best beam in the codebook.
Moreover, we also measure the beam pattern of the learned interference-unaware beam (plotted in \fref{exp-unaware:b}) in an anechoic chamber as shown in \fref{exp-unaware:c}.
After the interference-unaware beam is learned, the beam weights are saved and the interferer is turned on.
We then measure the signal and interference levels (with noise) of this learned interference-unaware beam. It is worth mentioning that the interference levels also depend on the position of the interferer.
In our experiments, we select $3$ different interferer positions. The measurement results of the interference-unaware beam with the different interferer placements are summarized in \tabref{Measure-Table}.

The performance of the interference-aware beam pattern learning algorithm is then benchmarked with that of the interference-unaware beam.
Before we delve into the detailed measurement results, it is worth pointing out that, since both the transmitter and the interferer have a direct LOS connection with the receiver, and given the channel characteristics in the mmWave bands, it is reasonable to assume that the existing LOS link has stronger power than the other NLOS links (if any) and hence is the dominant factor in the learning process.
Furthermore, due to the finite number of antennas at the receiver (which leads to the limited angular resolution), intuitively speaking, what really matters is the angle difference between the transmitter-receiver LOS link and the interferer-receiver LOS link. And in general, the ``closer'' the transmitter and the interferer are, the harder for the receiver to distinguish between them.
Specifically, we classify whether a pair of positions (of transmitter and interferer) being well-separated or not by comparing the angle difference with respect to the receiver with the half-power beam-width (HPBW). For a  ULA with half-wavelength antenna spacing, the HPBW can be approximately calculated as $\frac{1.78}{N}~\mathrm{rad}$ with $N$ being the number of antenna elements \cite{Richards2014}. Therefore, the HPBW of our adopted 16-antenna phased array is around $6.37^\circ$.
Next, based on the angular separation of the transmitter and the interferer of the considered scenarios (i.e., \fref{outdoor-exp}), we divide the measurement results into two categories and discuss them respectively.


\begin{figure*}[t]
	\centering
    \subfigure[EXP 1: Power Measurement]{ \includegraphics[width=0.34\textwidth]{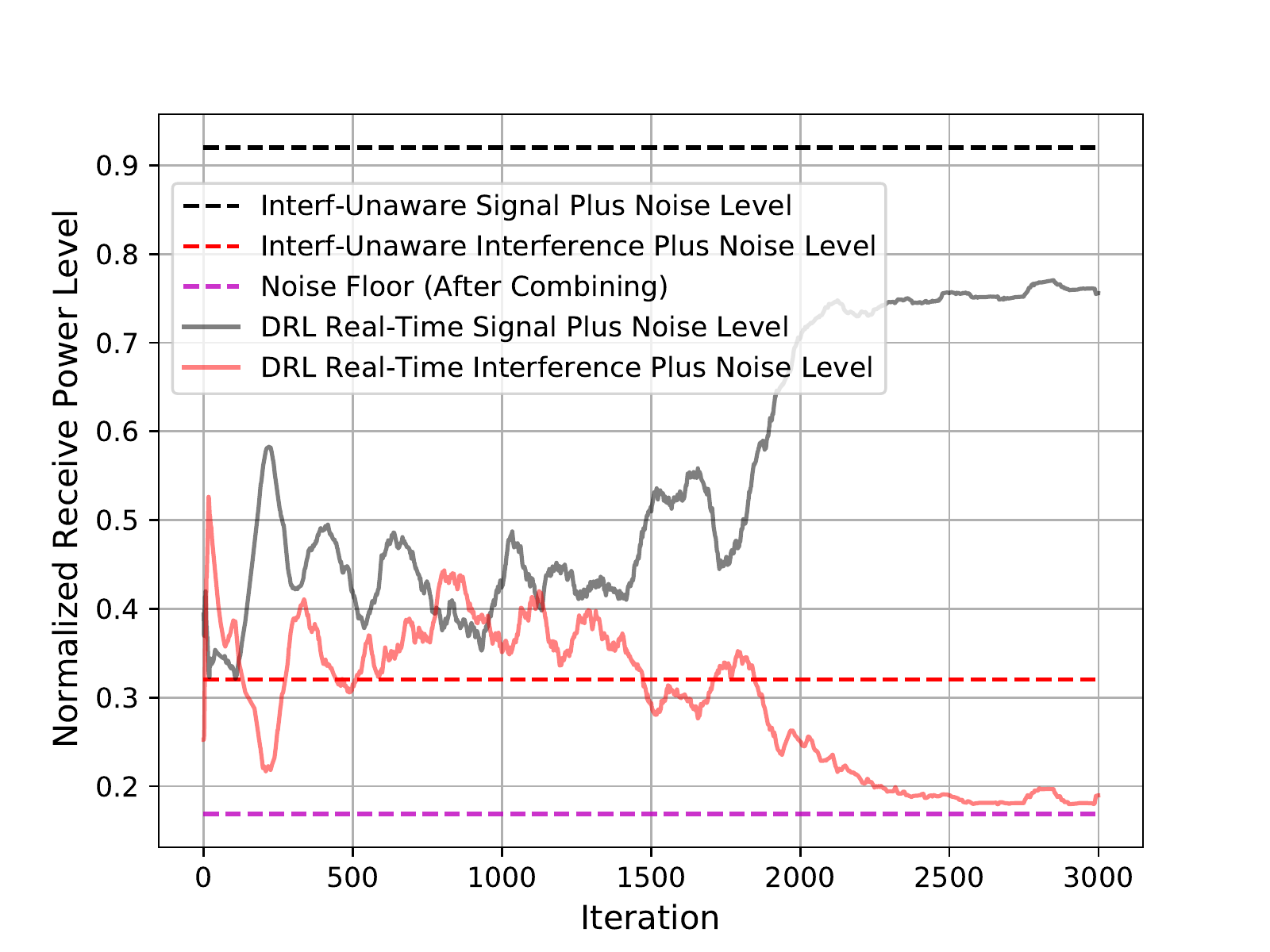} \label{exp_4:a} }
	\subfigure[EXP 1: SIR and INR Performance]{ \includegraphics[width=0.34\textwidth]{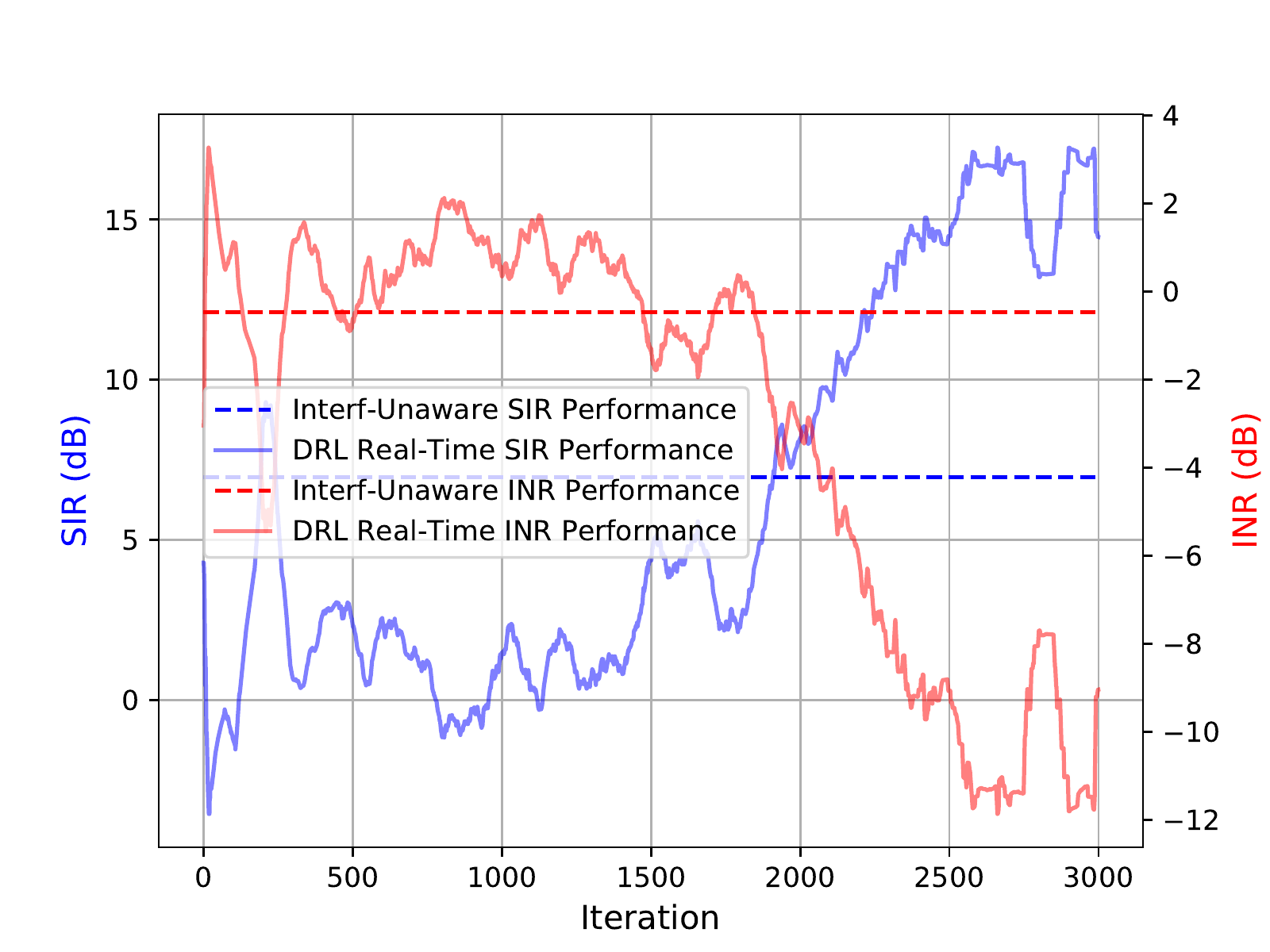} \label{exp_4:b} }
    \subfigure[EXP 1: Beam Pattern]{ \includegraphics[width=0.24\textwidth]{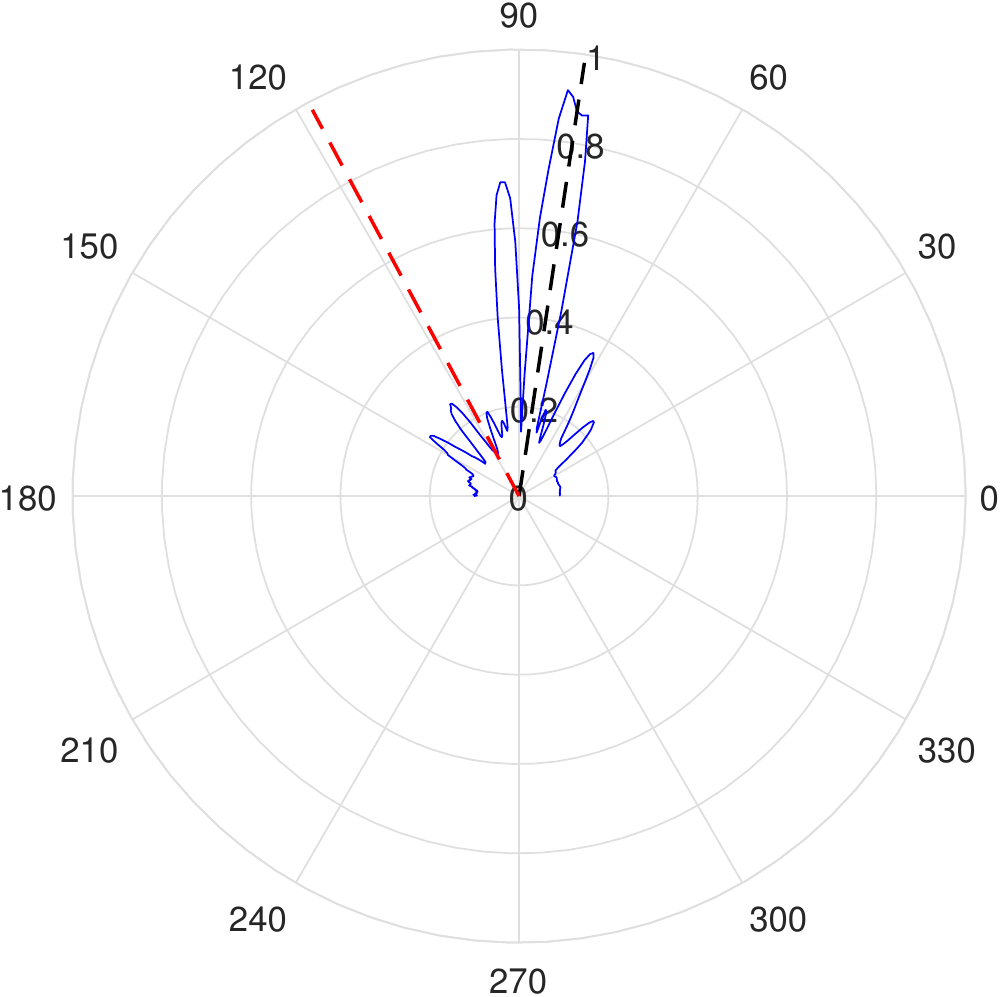} \label{exp_4:c} }

    \subfigure[EXP 2: Power Measurement]{ \includegraphics[width=0.34\textwidth]{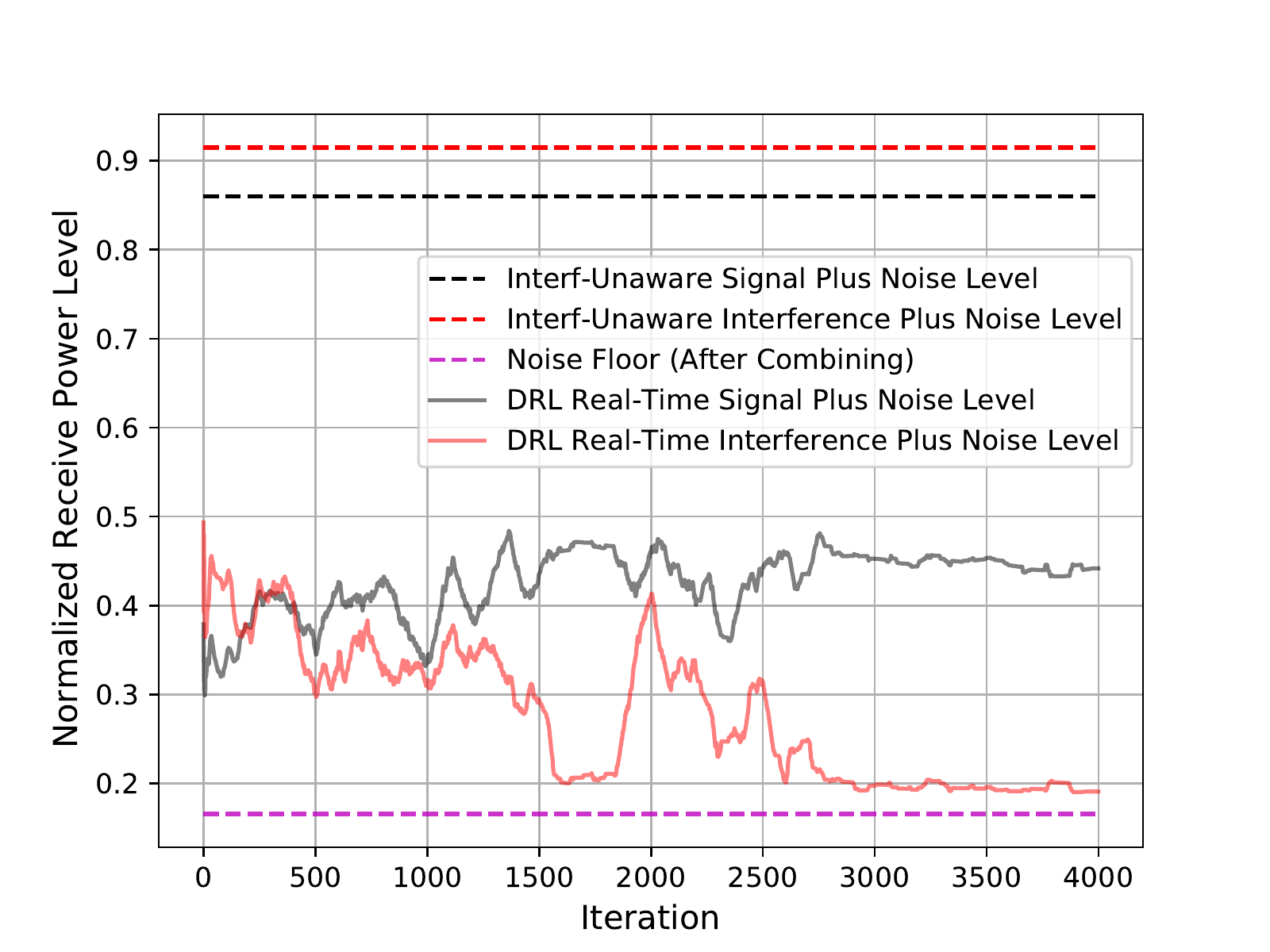} \label{exp_2:a} }
	\subfigure[EXP 2: SIR and INR Performance]{ \includegraphics[width=0.34\textwidth]{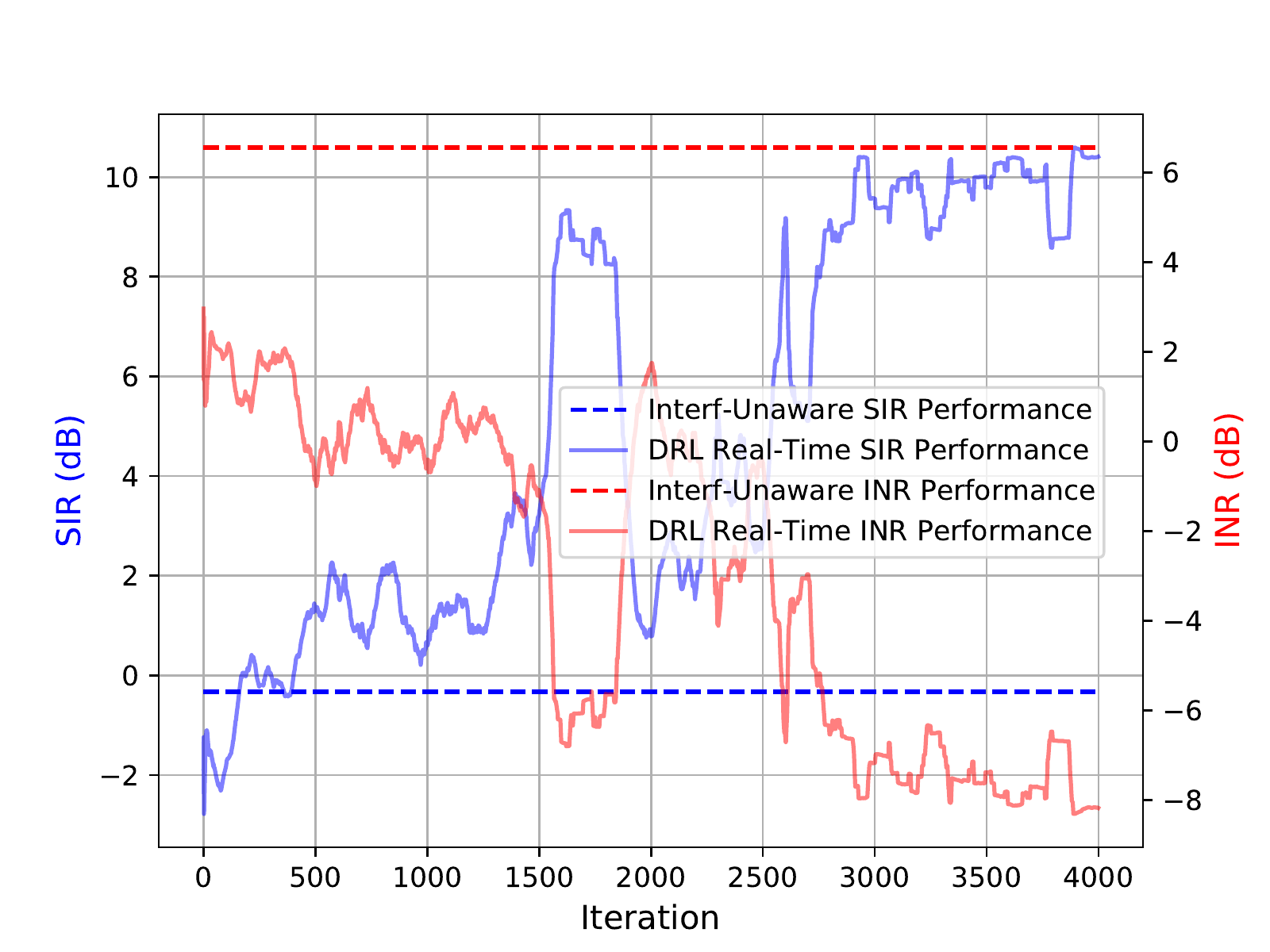} \label{exp_2:b} }
    \subfigure[EXP 2: Beam Pattern]{ \includegraphics[width=0.24\textwidth]{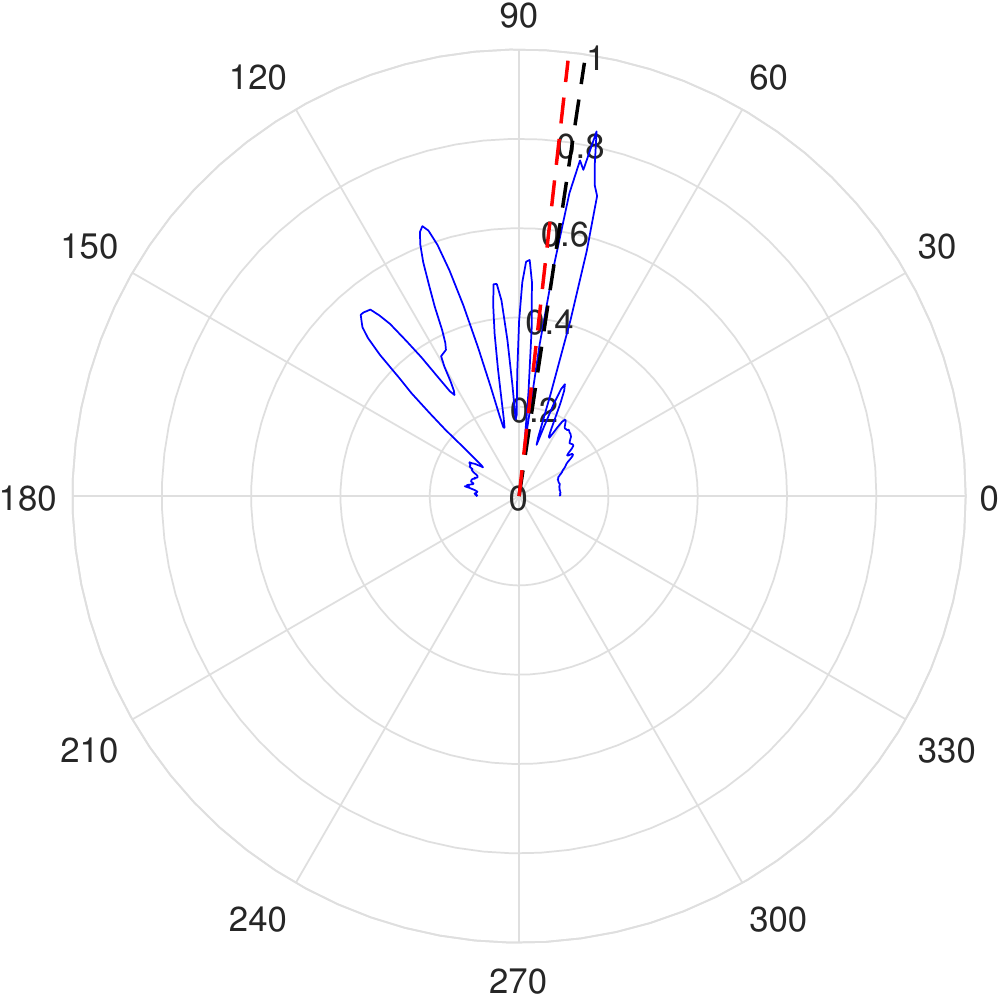} \label{exp_2:c} }

    \subfigure[EXP 3: Power Measurement]{ \includegraphics[width=0.34\textwidth]{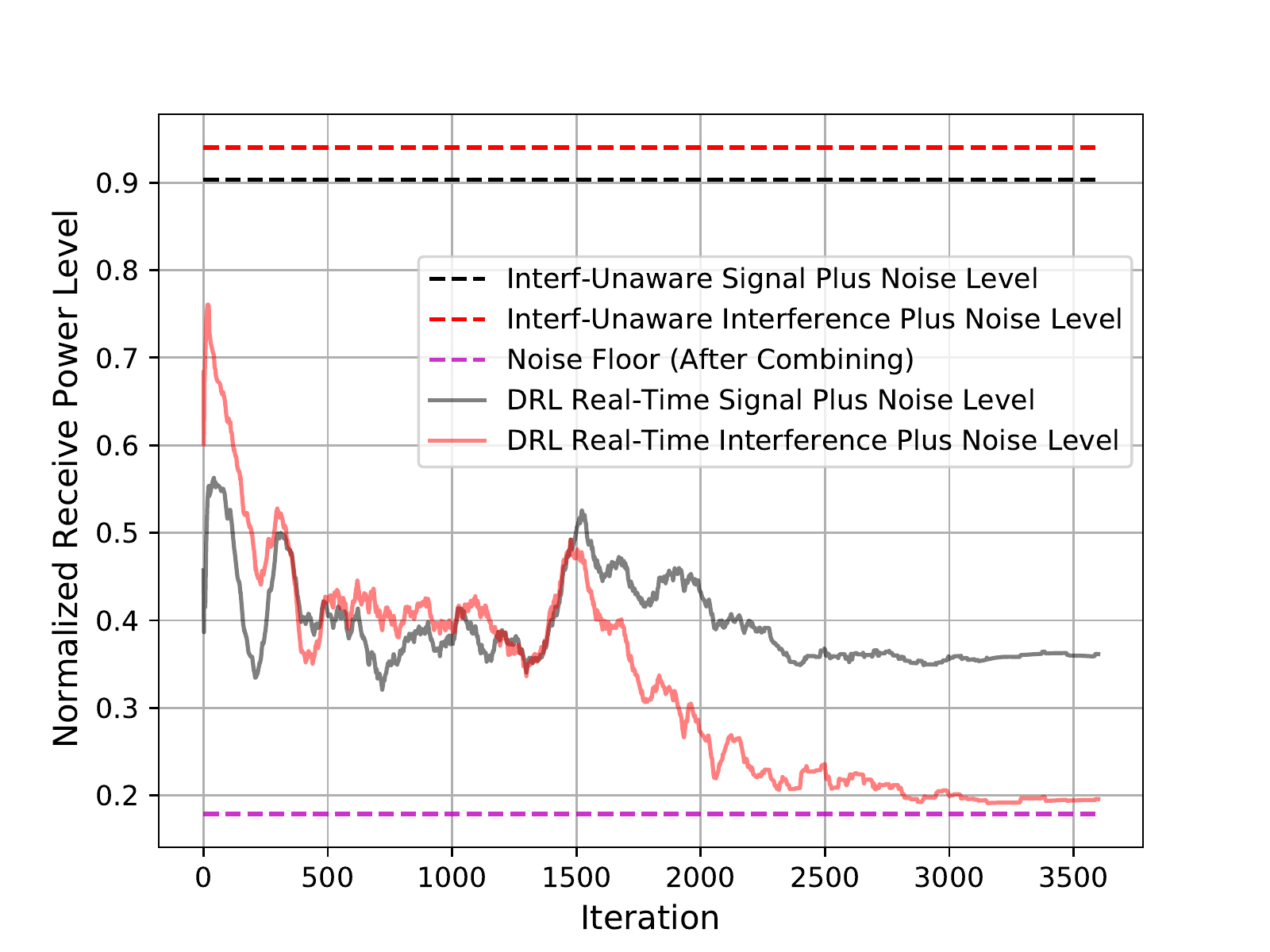} \label{exp_3:a} }
	\subfigure[EXP 3: SIR and INR Performance]{ \includegraphics[width=0.34\textwidth]{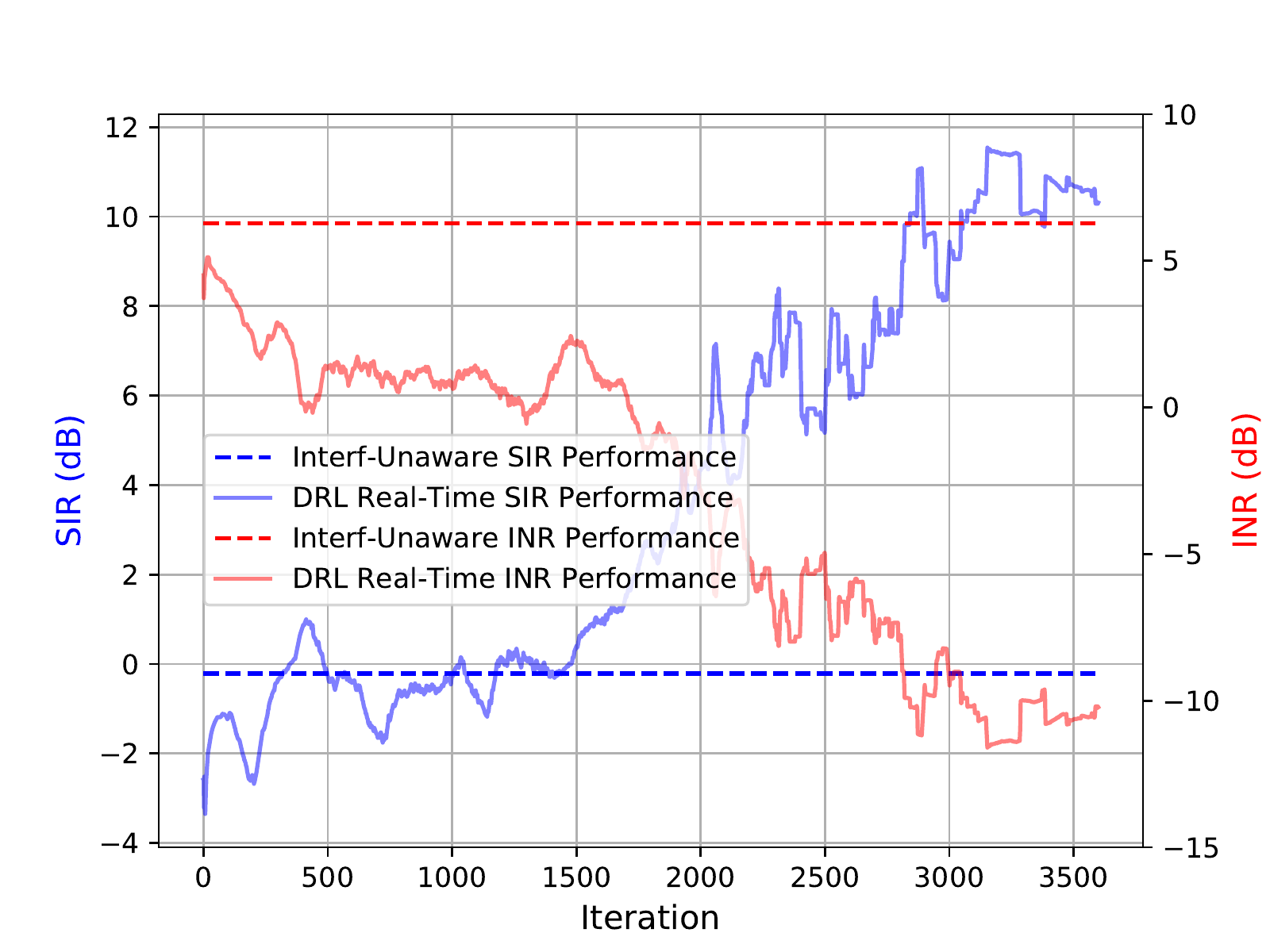} \label{exp_3:b} }
    \subfigure[EXP 3: Beam Pattern]{ \includegraphics[width=0.24\textwidth]{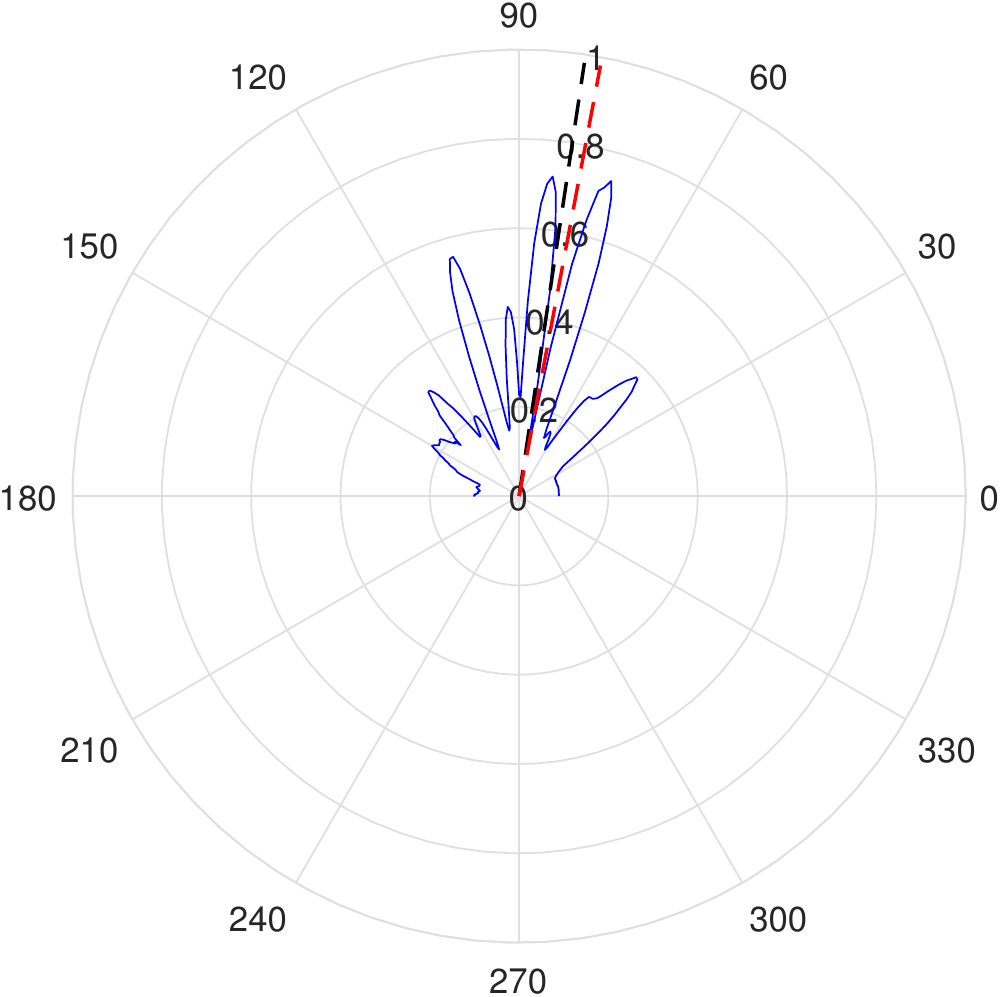} \label{exp_3:c} }
	\caption{The measurement results of the three experiments illustrated in \fref{outdoor-exp}, where the first column of figures shows the real-time receive power measurements and the second column of figures shows the corresponding SIR and INR performance. All these results are processed with a moving average of $100$ samples to smooth out the effect of noise. Finally, the third column of figures shows the learned interference-aware beam patterns with the black dashed line representing the direction of the desired signal and the red dashed line representing the direction of the interfering source.}
	\label{exp_123}
\end{figure*}


\subsubsection{When the transmitter and the interferer are well-separated}

We first study the case when the transmitter and the interferer are relatively well-separated, i.e., the angular separation is greater than that of the HPBW, in the experiment 1 (EXP 1 in \fref{outdoor-exp}).
In \fref{exp_4:a} and \fref{exp_4:b}, we plot the learning process of the experiment 1, where the angular separation of the transmitter and the interferer is around $40.33^\circ$. As can be seen in the figure, the performance of the interference-unaware beam is actually quite decent, yielding a SIR of $6.96$ dB and a INR of $-0.47$ dB, thanks to the significant angular separation.
However, it still introduces a certain level of interference which makes it comparable to the noise level and raises the interference plus noise level noticeably above the noise floor.
By contrast, the learned interference-aware beam is able to further suppress the interference to a great extent, making the INR even below $-10$ dB, i.e., \textbf{achieving a nearly 10 dB gain in INR}, while only sacrificing around $10\%$ of the desired signal power.
It is also worth mentioning that such performance is achieved \textbf{with only $3,500$ iterations and without knowing the channels of both the desired transmitter and the interferer.} Such relaxation on the system operations (such as synchronization and channel estimation) makes the proposed solution implementation friendly in most of the practical systems.

\subsubsection{When the transmitter and the interferer are extremely close}

Next, we study the case when the transmitter and the interferer are extremely close, i.e., the angular separation is much smaller than the HPBW, in the experiment 2 and 3 (EXP 2 and 3 in \fref{outdoor-exp}).
The angular separations of the transmitter and the interferer in both experiments are around $2^\circ$, which is only one third of the HPBW of the adopted phased array receiver.
It turns out that the proposed algorithm is still quite capable of suppressing the interference level. As can be seen in \fref{exp_2:b} and \fref{exp_3:b}, the SIR in both experiments all finally reaches over $10$ dB, and the INR level is also reduced to $-8$ dB and $-10$ dB respectively, \textbf{achieving almost the similar performance when the transmitter and the interferer are well-separated.}
However, different from the previous case, such great SIR and INR performances are traded with the significant sacrifices of the desired signal power. As indicated in \fref{exp_2:a} and \fref{exp_3:a}, the signal power is only around $50\%$ of that achieved in experiment 1, for instance. This also implies that \textbf{when the directions of the signal and interference are well aligned, the system normally needs to strike a delicate balance between SIR and SNR performances, in order to yield a meaningful SINR value.}
Such observation is also empirically confirmed by the measured learned beam patterns. As can be seen from \fref{exp_2:c} and \fref{exp_3:c}, the receiver intelligently shapes deep nulls towards the directions of the interference, which explains the achieved well interference suppression capability. However, as a compromise, the main-lobes of the beams are no longer pointing towards the desired transmitter, leaving only the side of the main-lobes leveraged to serve the target transmitter. This makes the receive signal power much weaker than that of the interference-unaware beam.
In summary, the real-world prototype confirms the effectiveness and robustness of the proposed solution in learning interference nulling beam patterns based solely on the power measurements. It also shows the promising gains brought by the intelligent online beam learning solution in realistic scenarios when compared with the off-the-shelf beams.

\section{Conclusion and Future Work} \label{sec:Con}

In this paper, we developed a sample-efficient online reinforcement learning based approach that can efficiently learn interference-aware beams. The proposed solution learns how to design beam patterns that can effectively manage interference, relying only on the power measurements and without any channel knowledge. This solution also relaxes the coherence/synchronization requirements of the system and respects the key hardware constraints of practical mmWave transceiver architectures. The results show that the proposed solution is capable of shaping nulls towards the interfering directions while maximizing the reception quality of the desired signal. When tested on a hardware proof-of-concept prototype based on real-world measurements, the proposed  interference-aware beam learning framework also demonstrating efficient beam pattern optimization performance.  Specifically, the developed solution was shown to improve the SNR and INR performance by at least 10 dB compared to the interference-unaware beams in all the tested scenarios. This is particularly important when the interferer is close to the transmitter. These SNR/INR gains can be translated to more than double the data rate in the considered scenarios. For the future work, it is important to extend the proposed framework to network settings with multiple basestation and decentralized beam learning capabilities.  It is also interesting to investigate the application of the developed beam learning approaches to full-duplex networks and cell-free massive MIMO architectures.

\linespread{1.2}


\begin{thebibliography}{10}
\providecommand{\url}[1]{#1}
\csname url@samestyle\endcsname
\providecommand{\newblock}{\relax}
\providecommand{\bibinfo}[2]{#2}
\providecommand{\BIBentrySTDinterwordspacing}{\spaceskip=0pt\relax}
\providecommand{\BIBentryALTinterwordstretchfactor}{4}
\providecommand{\BIBentryALTinterwordspacing}{\spaceskip=\fontdimen2\font plus
\BIBentryALTinterwordstretchfactor\fontdimen3\font minus
  \fontdimen4\font\relax}
\providecommand{\BIBforeignlanguage}[2]{{%
\expandafter\ifx\csname l@#1\endcsname\relax
\typeout{** WARNING: IEEEtran.bst: No hyphenation pattern has been}%
\typeout{** loaded for the language `#1'. Using the pattern for}%
\typeout{** the default language instead.}%
\else
\language=\csname l@#1\endcsname
\fi
#2}}
\providecommand{\BIBdecl}{\relax}
\BIBdecl

\bibitem{Zhang2022Online}
Y.~Zhang and A.~Alkhateeb, ``{Online Beam Learning for Interference Nulling in
  Hardware-Constrained mmWave MIMO Systems},'' in \emph{{2022 56th Asilomar
  Conference on Signals, Systems, and Computers}}, 2022.

\bibitem{Alkhateeb2014MIMO}
A.~{Alkhateeb}, J.~{Mo}, N.~{Gonzalez-Prelcic}, and R.~W. {Heath}, ``{MIMO
  Precoding and Combining Solutions for Millimeter-Wave Systems},'' \emph{IEEE
  Communications Magazine}, vol.~52, no.~12, pp. 122--131, 2014.

\bibitem{Heath2016}
R.~W. {Heath}, N.~{González-Prelcic}, S.~{Rangan}, W.~{Roh}, and A.~M.
  {Sayeed}, ``{An Overview of Signal Processing Techniques for Millimeter Wave
  MIMO Systems},'' \emph{IEEE Journal of Selected Topics in Signal Processing},
  vol.~10, no.~3, pp. 436--453, 2016.

\bibitem{Molisch2017}
A.~F. Molisch, V.~V. Ratnam, S.~Han, Z.~Li, S.~L.~H. Nguyen, L.~Li, and
  K.~Haneda, ``{Hybrid Beamforming for Massive MIMO: A Survey},'' \emph{IEEE
  Communications Magazine}, vol.~55, no.~9, pp. 134--141, 2017.

\bibitem{Zhang2022Reinforcement}
Y.~Zhang, M.~Alrabeiah, and A.~Alkhateeb, ``{Reinforcement Learning of Beam
  Codebooks in Millimeter Wave and Terahertz MIMO Systems},'' \emph{{IEEE
  Transactions on Communications}}, vol.~70, no.~2, pp. 904--919, 2022.

\bibitem{Alrabeiah2020Neural}
M.~Alrabeiah, Y.~Zhang, and A.~Alkhateeb, ``{Neural Networks Based Beam
  Codebooks: Learning mmWave Massive MIMO Beams That Adapt to Deployment and
  Hardware},'' \emph{{IEEE Transactions on Communications}}, vol.~70, no.~6,
  pp. 3818--3833, 2022.

\bibitem{Heng2021}
Y.~Heng, J.~G. Andrews, J.~Mo, V.~Va, A.~Ali, B.~L. Ng, and J.~C. Zhang, ``{Six
  Key Challenges for Beam Management in 5.5G and 6G Systems},'' \emph{IEEE
  Communications Magazine}, vol.~59, no.~7, pp. 74--79, 2021.

\bibitem{Venkatesan2007}
S.~Venkatesan, A.~Lozano, and R.~Valenzuela, ``{Network MIMO: Overcoming
  Intercell Interference in Indoor Wireless Systems},'' in \emph{{2007
  Conference Record of the Forty-First Asilomar Conference on Signals, Systems
  and Computers}}, 2007, pp. 83--87.

\bibitem{Yetis2010}
C.~M. Yetis, T.~Gou, S.~A. Jafar, and A.~H. Kayran, ``{On Feasibility of
  Interference Alignment in MIMO Interference Networks},'' \emph{{IEEE
  Transactions on Signal Processing}}, vol.~58, no.~9, pp. 4771--4782, 2010.

\bibitem{Choi2007}
W.~Choi and J.~G. Andrews, ``{Downlink performance and capacity of distributed
  antenna systems in a multicell environment},'' \emph{{IEEE Transactions on
  Wireless Communications}}, vol.~6, no.~1, pp. 69--73, 2007.

\bibitem{Gesbert2010}
D.~Gesbert, S.~Hanly, H.~Huang, S.~Shamai~Shitz, O.~Simeone, and W.~Yu,
  ``{Multi-Cell MIMO Cooperative Networks: A New Look at Interference},''
  \emph{{IEEE Journal on Selected Areas in Communications}}, vol.~28, no.~9,
  pp. 1380--1408, 2010.

\bibitem{Irmer2011}
R.~Irmer \emph{et~al.}, ``{Coordinated multipoint: Concepts, performance, and
  field trial results},'' \emph{{IEEE Communications Magazine}}, vol.~49,
  no.~2, pp. 102--111, 2011.

\bibitem{Ngo2017}
H.~Q. Ngo, A.~Ashikhmin, H.~Yang, E.~G. Larsson, and T.~L. Marzetta,
  ``{Cell-Free Massive MIMO Versus Small Cells},'' \emph{{IEEE Transactions on
  Wireless Communications}}, vol.~16, no.~3, pp. 1834--1850, 2017.

\bibitem{Interdonato2020}
G.~Interdonato, M.~Karlsson, E.~Björnson, and E.~G. Larsson, ``{Local Partial
  Zero-Forcing Precoding for Cell-Free Massive MIMO},'' \emph{{IEEE
  Transactions on Wireless Communications}}, vol.~19, no.~7, pp. 4758--4774,
  2020.

\bibitem{Demirhan2022}
U.~Demirhan and A.~Alkhateeb, ``Enabling cell-free massive mimo systems with
  wireless millimeter wave fronthaul,'' \emph{IEEE Transactions on Wireless
  Communications}, pp. 1--1, 2022.

\bibitem{Zhang2008Exploiting}
R.~Zhang and Y.-C. Liang, ``{Exploiting Multi-Antennas for Opportunistic
  Spectrum Sharing in Cognitive Radio Networks},'' \emph{{IEEE Journal of
  Selected Topics in Signal Processing}}, vol.~2, no.~1, pp. 88--102, 2008.

\bibitem{Li2016Optimum}
B.~Li, A.~P. Petropulu, and W.~Trappe, ``{Optimum Co-Design for Spectrum
  Sharing between Matrix Completion Based MIMO Radars and a MIMO Communication
  System},'' \emph{{IEEE Transactions on Signal Processing}}, vol.~64, no.~17,
  pp. 4562--4575, 2016.

\bibitem{Qian2018}
J.~Qian, M.~Lops, L.~Zheng, X.~Wang, and Z.~He, ``{Joint System Design for
  Coexistence of MIMO Radar and MIMO Communication},'' \emph{{IEEE Transactions
  on Signal Processing}}, vol.~66, no.~13, pp. 3504--3519, 2018.

\bibitem{Alkhateeb2014}
A.~Alkhateeb, O.~El~Ayach, G.~Leus, and R.~Heath, ``{Channel Estimation and
  Hybrid Precoding for Millimeter Wave Cellular Systems},'' \emph{IEEE Journal
  of Selected Topics in Signal Processing}, vol.~8, no.~5, pp. 831--846, Oct.
  2014.

\bibitem{Ayach2014}
O.~E. Ayach, S.~Rajagopal, S.~Abu-Surra, Z.~Pi, and R.~W. Heath, ``{Spatially
  Sparse Precoding in Millimeter Wave MIMO Systems},'' \emph{{IEEE Transactions
  on Wireless Communications}}, vol.~13, no.~3, pp. 1499--1513, 2014.

\bibitem{Yu2016}
X.~Yu, J.-C. Shen, J.~Zhang, and K.~B. Letaief, ``{Alternating Minimization
  Algorithms for Hybrid Precoding in Millimeter Wave MIMO Systems},''
  \emph{{IEEE Journal of Selected Topics in Signal Processing}}, vol.~10,
  no.~3, pp. 485--500, 2016.

\bibitem{Alkhateeb2015Limited}
{A. {Alkhateeb} and G. {Leus} and R. W. {Heath}}, ``{Limited Feedback Hybrid
  Precoding for Multi-User Millimeter Wave Systems},'' \emph{{IEEE Transactions
  on Wireless Communications}}, vol.~14, no.~11, pp. 6481--6494, 2015.

\bibitem{Sohrabi2016}
F.~Sohrabi and W.~Yu, ``{Hybrid Digital and Analog Beamforming Design for
  Large-Scale Antenna Arrays},'' \emph{{IEEE Journal of Selected Topics in
  Signal Processing}}, vol.~10, no.~3, pp. 501--513, 2016.

\bibitem{Zhan2021Interference}
J.~Zhan and X.~Dong, ``{Interference Cancellation Aided Hybrid Beamforming for
  mmWave Multi-User Massive MIMO Systems},'' \emph{{IEEE Transactions on
  Vehicular Technology}}, vol.~70, no.~3, pp. 2322--2336, 2021.

\bibitem{Satyanarayana2019}
K.~Satyanarayana, M.~El-Hajjar, P.-H. Kuo, A.~Mourad, and L.~Hanzo, ``{Hybrid
  Beamforming Design for Full-Duplex Millimeter Wave Communication},''
  \emph{{IEEE Transactions on Vehicular Technology}}, vol.~68, no.~2, pp.
  1394--1404, 2019.

\bibitem{Roberts2021}
I.~P. Roberts, J.~G. Andrews, and S.~Vishwanath, ``{Hybrid Beamforming for
  Millimeter Wave Full-Duplex Under Limited Receive Dynamic Range},''
  \emph{{IEEE Transactions on Wireless Communications}}, vol.~20, no.~12, pp.
  7758--7772, 2021.

\bibitem{Zhu2020}
L.~Zhu, J.~Zhang, Z.~Xiao, X.~Cao, X.-G. Xia, and R.~Schober,
  ``{Millimeter-Wave Full-Duplex UAV Relay: Joint Positioning, Beamforming, and
  Power Control},'' \emph{{IEEE Journal on Selected Areas in Communications}},
  vol.~38, no.~9, pp. 2057--2073, 2020.

\bibitem{Zhang2019On}
Y.~Zhang, M.~Xiao, S.~Han, M.~Skoglund, and W.~Meng, ``{On Precoding and Energy
  Efficiency of Full-Duplex Millimeter-Wave Relays},'' \emph{{IEEE Transactions
  on Wireless Communications}}, vol.~18, no.~3, pp. 1943--1956, 2019.

\bibitem{Lorenz2005}
R.~Lorenz and S.~Boyd, ``{Robust minimum variance beamforming},'' \emph{{IEEE
  Transactions on Signal Processing}}, vol.~53, no.~5, pp. 1684--1696, 2005.

\bibitem{Dahrouj2010}
H.~Dahrouj and W.~Yu, ``{Coordinated beamforming for the multicell
  multi-antenna wireless system},'' \emph{{IEEE Transactions on Wireless
  Communications}}, vol.~9, no.~5, pp. 1748--1759, 2010.

\bibitem{Alkhateeb2016Frequency}
A.~Alkhateeb and R.~W. Heath, ``Frequency selective hybrid precoding for
  limited feedback millimeter wave systems,'' \emph{IEEE Transactions on
  Communications}, vol.~64, no.~5, pp. 1801--1818, 2016.

\bibitem{3gpp.38.802}
3GPP, ``{Study on new radio access technology: Physical layer aspects},'' Tech.
  Rep. 38.802, 2017, version 14.2.2.

\bibitem{NNUnivApprox}
K.~Hornik, M.~Stinchcombe, and H.~White, ``Multilayer feedforward networks are
  universal approximators,'' \emph{Neural networks}, vol.~2, no.~5, pp.
  359--366, 1989.

\bibitem{Richards2014}
M.~A. Richards, \emph{{Fundamentals of radar signal processing}}.\hskip 1em
  plus 0.5em minus 0.4em\relax {McGraw-Hill Education}, 2014.

\end{thebibliography}
\end{document}